\documentclass[useAMS,usenatbib]{mn2e}
\usepackage{graphicx}
\usepackage{caption}
\usepackage{subcaption}
\usepackage{natbib}
\usepackage{amsmath}
\usepackage{float}
\usepackage{pinlabel}
\usepackage{placeins}
\usepackage{rotating}
\usepackage{color}
\bibpunct{(}{)}{;}{a}{}{,}

\newcommand{\ie}{$i.e.,\;$}
\newcommand{\eg}{$e.g.,\;$}

\usepackage{txfonts}

\title[]{Discovery of rare double$-$lobe radio galaxies hosted in spiral galaxies}
\author[Singh et al.]{Veeresh Singh$^{1}$\thanks{E-mail: Singhv4@ukzn.ac.za}, 
C. H. Ishwara-Chandra$^{2}$, Jonathan Sievers$^{1}$, Yogesh Wadadekar$^{2}$,  
\newauthor Matt Hilton$^{3}$ and Alexandre Beelen$^{4}$ 
\\
$^{1}$Astrophysics and Cosmology Research Unit, School of Chemistry and Physics, University of KwaZulu-Natal, Durban 4041, South Africa \\
$^{2}$National Centre for Radio Astrophysics, TIFR, Post Bag 3, Ganeshkhind, Pune 411007, India\\ 
$^{3}$Astrophysics and Cosmology Research Unit, School of Mathematics, Statistics \& Computer Science, University of KwaZulu-Natal, Durban 4041, South Africa \\
$^{4}$Institut d'Astrophysique Spatiale, B$\hat{\rm a}$t. 121, Universit{\'e} Paris-Sud, 91405 Orsay Cedex, France \\
}

\begin{document}

\date{Accepted ........; Received .........; in original form ........}

\pagerange{\pageref{firstpage}--\pageref{lastpage}} \pubyear{2015}

\maketitle

\label{firstpage}

\begin{abstract}
Double-lobe radio galaxies in the local Universe have traditionally been found to be hosted in elliptical or lenticular galaxies.   
We report the discovery of four spiral-host double-lobe radio galaxies (J0836+0532, J1159+5820, J1352+3126 and J1649+2635) 
that are discovered by cross-matching a large sample of 187005 spiral galaxies from SDSS DR7 to 
the full catalogues of FIRST and NVSS. J0836+0532 is reported for the first time. 
The host galaxies are forming stars at an average rate of $1.7$--$10$\,M$_{\odot}$\,yr$^{-1}$ and 
possess Super Massive Black Holes (SMBHs) with masses of a few times 10$^{8}$ M$_{\odot}$. 
Their radio morphologies are similar to FR-II radio galaxies with total projected linear sizes ranging from 
86\,kpc to 420\,kpc, but their total 1.4\,GHz radio luminosities are only in the range 10$^{24}$ $-$ 10$^{25}$\,W\,Hz$^{-1}$. 
We propose that the formation of spiral-host double-lobe radio galaxies can be attributed to more than one factor, 
such as the occurrence of strong interactions, mergers, and the presence of unusually massive SMBHs, such that the 
spiral structures are not destroyed. 
Only one of our sources (J1649+2635) is found in a cluster environment, 
indicating that processes other than accretion through cooling flows {\eg}galaxy-galaxy mergers or interactions could be 
plausible scenarios for triggering radio-loud AGN activity in spiral galaxies.
   
\end{abstract}


\section{Introduction}
Radio galaxies in the local Universe are found to be exclusively hosted in massive, gas poor elliptical galaxies characterised by 
feeble star formation rates \citep{Urry95,Wilson95,McLure04b,Best05}. 
It is believed that the relativistic jets emanating from the accreting Super Massive Black Holes (SMBH) at their centres can easily plough through
the rarer interstellar medium (ISM) of elliptical galaxies and reach scales of 100--1000 kpc.   
In contrast, the denser ISM present in spiral galaxies impedes jets and confines them to within the host galaxy as often seen in 
Seyfert galaxies \citep{Gallimore06,Schawinski11}. 
Arguably, the nature of the host galaxy ISM is one of the key parameters responsible for the form of the AGN jets that come out of the galaxy \citep{Capetti06}. 
Moreover, other factors such as the mass and spin of the SMBH are also believed to play a crucial role in the formation of larger jets in radio galaxies 
\citep{Laor2000,Liu06,Chiaberge11}. \par 
It has been shown that active galactic nuclei (AGN) release vast amounts of energy into the ISM via feedback processes that cause an increase 
in the temperature of the ISM gas, expel it via outflows, and quench star formation \citep[see][]{Fabian12}.
AGN feedback can be either via AGN radiation ({\ie}`quasar mode'), or through the interaction of the jet with the ISM 
(known as kinetic or `radio mode'). 
Radio galaxies with extremely large and fast jets are able to inject enormous amounts of energy into the ISM 
and even into the intergalactic medium (IGM; \cite{McNamara07}). 
In general, the star formation efficiency of radio galaxies in the local Universe is found to be $\sim$ 10--50 times less 
than that in normal galaxies, and this fact can be reconciled via AGN feedback \citep{Nesvadba10}. 
However, it has also been shown that not all radio galaxies have low star formation efficiency, and so the impact of feedback on a galaxy 
may evolve \citep[{\eg}][]{Labiano13,Labiano14}. 
Therefore, the study of radio galaxies hosted in spiral, disk-dominated galaxies with high star formation efficiency is 
key to understanding how AGN feedback affects the properties of their hosts.
In fact, such studies are important for an overall understanding of galaxy evolution, as 
AGN feedback is believed to be responsible for several observed phenomena  
{\eg}the correlation between black hole and host galaxy bulge mass \citep{Magorrian98,Tremaine02}, and
the fast transition of early-type galaxies from the blue cloud to the red sequence \citep{Schawinski07,Kaviraj11}. 
\par
Furthermore, despite the presence of vast numbers of radio and optical observations, 
powerful radio galaxies hosted in spiral disk-dominated galaxies are extremely rare. 
Hitherto, only a handful of such sources have been found 
{\eg}0313-192 \citep{Ledlow98}; Speca \citep{Hota11}; J2345-0449 \citep{Bagchi14} and J1649+2635 \citep{Mao15}.
In this paper we present a sample of four radio galaxies hosted in spiral galaxies. 
These galaxies are discovered by cross-matching a large sample of 187005 spiral galaxies derived 
from Sloan Digital Sky Survey (SDSS\footnote{http://www.sdss.org/}; \cite{Abazajian09}) 
Data Release 7 (DR7) with the full Faint Images of the Radio Sky at Twenty-cm (FIRST\footnote{http://sundog.stsci.edu/}; \cite{Becker95}) and 
the NRAO VLA Sky Survey (NVSS\footnote{http://www.cv.nrao.edu/nvss/}; \cite{Condon98}) radio data.
Our study is the first systematic attempt to search for spiral-host double-lobe radio galaxies using large area optical and radio surveys. 
\\
This paper is structured as follows. 
In Section 2, we describe the data used to search for spiral-host double-lobe radio galaxies. 
Our search strategy is discussed in Section 3. 
In Section 4, we describe the methods used to estimate various properties ({\eg}star formation rates, black holes masses, large-scale 
environments) of the host galaxies. The properties of individual sources are discussed in Section 5. We discuss our findings in
Section 6 and present our conclusions in Section 7. In this paper we assume a cosmology with 
$H_{0} = 71$\,km\,s$^{-1}$\,Mpc$^{-1}$, ${\Omega}_{\rm M} = 0.27$, and ${\Omega}_{\Lambda} = 0.73$.

\section{Data}
\subsection{Optical}

We use a sample of 187005 spiral galaxies from the \cite{Meert15} catalogue that provides morphological classification of galaxies of 
the SDSS DR7 spectroscopic sample \citep{Abazajian09}. 
The galaxy morphology ({\ie}spiral, elliptical or uncertain) is determined quantitatively by an automated two-dimensional photometric 
decomposition of the $r$-band surface-brightness using the \textsc{Pymorph} software pipeline \citep{Vikram10} and 
\textsc{Galfit} \citep{Peng02}. 
The \cite{Meert15} catalogue is one of the most recent and extensive works to classify the morphologies of SDSS galaxies. 
This catalogue has substantial overlap with other galaxy morphology catalogues based on full or partial SDSS data 
\citep[{\eg}][]{Simard11,Kelvin12,Lackner12}, and the `Galaxy Zoo' citizen scientist based programme \citep{Lintott08,Lintott11}. 
A detailed comparison of these various catalogues is given in \cite{Meert15}.  
\subsection{Radio}
To find the radio counterparts of the spiral galaxies in our sample,  we use 
the FIRST survey that was carried out at 1.4 GHz with the Very Large Array (VLA) in its `B' configuration, and 
covers a sky region over 10,000 square degrees around the North and South Galactic Caps \citep{Becker95}.
The FIRST radio maps have a resolution of $\sim$ 5$\arcsec$ and a typical noise rms of $\sim 0.15$\, mJy/beam. 
We have used the most recent version of FIRST source catalogue (released in March 2014), that lists peak as well as integrated flux densities 
and radio sizes derived by fitting a two-dimensional Gaussian to a source detected at a flux limit of $\sim$ 1\,mJy (SNR $\geq$ 5). 
The astrometric reference frame of the FIRST maps is accurate to $\sim$ 0.05$\arcsec$ and individual sources have $\sim$ 90$\%$ confidence 
error circles of radius $<$ 0.5$\arcsec$ at the 3\,mJy level and $\sim$ 1$\arcsec$.0 at the 1\,mJy survey threshold \citep[see][]{Becker95}.\\
The relatively high resolution (smaller synthesized beamsize) FIRST observations can miss extended 
low-surface-brightness emission as the flux density per beam falls below the sensitivity limit ({\ie}it is resolved out), and so
we use NVSS radio data to detect fainter, extended diffuse emission from the lobes of radio galaxies. 
NVSS is a 1.4\,GHz continuum survey carried out by the VLA in its `D' configuration. 
It covers the entire sky north of $-40\deg$ declination and provides radio 
images with $\sim 45\arcsec$ resolution and sensitivity $\sim 2.5$\,mJy at the 5$\sigma$ level \citep{Condon98}.
We use the NVSS catalogue which contains over 1.8 million unique detections brighter than 2.5\,mJy 
and gives total integrated 1.4\,GHz flux densities for all the radio sources. 
The astrometric accuracy ranges from $\sim 1\arcsec$.0 for the brightest NVSS detections 
to about 7$\arcsec$.0 for the faintest detections.\\
Some studies on the search for radio counterparts of spiral galaxies
used the Unified Radio Catalogue \citep{Kimball08}, which contains data from
both the FIRST and NVSS surveys, and also lists sources detected in the Green
Bank 6\,cm survey (GB6) and the 92\,cm Westerbork Northern Sky Survey (WENSS). 
We caution that the unified radio catalogue is a mere
compilation of the closest NVSS sources to a FIRST source within a circle
of 30$\arcsec$ radius and vice-versa for a NVSS source. Users are
advised to opt for their own criteria to use true counterparts
\citep[see][]{Kimball08}. Given these limitations, we choose 
to use the FIRST and NVSS catalogues to find radio counterparts of SDSS spiral galaxies.

\section{Search strategy}
To search for double-lobe radio galaxies associated with SDSS DR 7 spiral galaxies we make optimal use of both FIRST and NVSS 
in a multi-stage process. 
Both FIRST and NVSS are carried out at 1.4\,GHz, but with different resolutions and sensitivities, which make them complementary to each other. 
The relatively high resolution ($\sim 5\arcsec$ beam) and sensitive (5$\sigma$ rms $\sim 1$\,mJy/beam) FIRST observations 
can efficiently detect faint, kpc-scale radio structures that may remain unresolved or undetected in the NVSS. 
On the other hand, NVSS observations of relatively low resolution ($\sim 45\arcsec$ beam, 5$\sigma$ rms $\sim 2.5$\,mJy/beam) can 
efficiently detect diffuse, low-surface-brightness, extended radio emission 
from hundreds-of-kpc-scale large radio lobes that may get resolved out in FIRST.
Our multi-stage search process involves the following steps. \\
(i){\it Using only FIRST data} : 
In general, radio galaxy morphology consists of a central AGN core emission component and a pair of bipolar jets terminating in lobes. 
For the purpose of cross-matching radio sources to their optical counterparts, we refer to the unresolved central radio-emitting component 
as `the core', although it may contain emission from parts of the jets close to the AGN core.    
To find double-lobe radio sources hosted in spiral galaxies, 
we first look for sources in which the core component is closely matched to an optical source within 
3$\arcsec$, and the two radio lobes are clearly detected within a circular region of radius 180$^{\arcsec}$. 
Double-lobe radio sources larger than 180$^{\arcsec}$ can be more effectively detected in the NVSS and are searched for in step (ii).       
\par 
In cross-matching optical and radio sources, the random offset between radio and optical 
positions is mainly due to positional uncertainties in the FIRST and SDSS sources. 
The systematic offset between FIRST and SDSS spiral galaxies positions is negligible 
($\Delta$RA = (RA$_{\rm SDSS}$ - RA$_{\rm FIRST}$) $\sim$ -0$^{\arcsec}$.087 
and $\Delta$DEC = (DEC$_{\rm SDSS}$ - DEC$_{\rm FIRST}$) $\sim$ -0$^{\arcsec}$.0006) as 
compared to the positional uncertainties. 
The choice of search radius in cross-matching is a trade off between the completeness and contamination, 
{\ie} a smaller matching radius reduces the contamination at the expense of completeness. 
For example, using a search radius of 1$^{\arcsec}$.0 - 2$^{\arcsec}$.0 for matching FIRST and SDSS sources, 
greatly reduces the contamination, but introduces bias against true radio counterparts mainly at the fainter end, 
where the astrometric uncertainties are larger \citep[{see}][]{Lu07,McGreer09}. 
Also, for extended radio sources the fitted radio centroid position may not correspond directly 
to the optical position.
We choose a search radius of 3$\arcsec$.0 owing to the fact that the distribution of the separation between the 
optical and radio positions in the cross-matched sample of FIRST and SDSS galaxies is approximately Gaussian up to 
3$\arcsec$.0, and tails off beyond it with an increasing number of 
contaminants, which is consistent with previous studies \citep[{see}][]{Wadadekar04,deVries06,Singh15}.
\par
We cross-match 187005 SDSS DR7 spiral galaxies to the FIRST catalogue using a search radius of 3${\arcsec}$, 
and this simple one-to-one cross-matching yields a total of 5030 sources. 
To find double-lobe radio sources among these 5030 galaxies, we search for the presence of multicomponent radio sources in 
a circle of radius 180$\arcsec$, centred at the optical position. 
We find that there are a total of 953 sources with two or more extended radio components within this matching radius. 
Visual inspection of all 953 sources using 6$\arcmin$~$\times$~6$\arcmin$ FIRST image cutouts shows 
that in the majority of cases, the extended components are completely unrelated to the central component, and 
in several cases, the central component matching with the spiral galaxy is found to be a radio lobe of an offset radio galaxy. 
We excluded all such cases. 
Since our objective is to search for sources with distinct double-lobe radio morphology, we also 
excluded sources in which FIRST radio morphology appears somewhat extended with no apparent 
double-lobe radio structure and radio size resides within the optical size of the host galaxy 
({\ie}possible cases of Seyfert galaxies or compact radio sources). 
Using FIRST data we finally obtain a total of four sources ({\ie}J1008+5026, J1159+5820, J1450-0106 and J1649+2635) exhibiting 
distinct double-lobes and a central core component associated with a spiral galaxy. 
\\
(ii) {\it Using NVSS and FIRST data} : 
In order to detect radio galaxies with hundreds-of-kpc to Mpc-scale jet-lobe structures, 
we search for the presence of extended radio components detected in the NVSS in a circle of radius 600$\arcsec$, centred 
at the optical positions of 5030 galaxies that possess a central radio core component 
detected in the FIRST.
There are a total of 498 sources having one or more extended ($>45\arcsec$) components with high 
signal-to-noise ratio (S/N $\geq$ 10). 
We visually inspected all these 498 sources using 30$\arcmin$ $\times$ 30$\arcmin$ NVSS image cut-outs and 
found that in the majority of cases, the extended radio components are completely unrelated to the central 
component. 
In several cases, the NVSS radio source appears extended due to closely placed more than one unrelated radio 
sources that are clearly seen in the corresponding FIRST images. 
We finally find only four sources ({\ie}J0836+0532, J1159+5820, J1352+3126 and J1649+2635) exhibiting 
double-lobe radio morphology. 
Two (J1159+5820 and J1649+2635) of these four sources were already found using only FIRST data in step (i). 
However, NVSS shows an additional pair of lobes in J1159+5820 that are completely undetected in FIRST 
due to the inherent limited sensitivity of FIRST to detect low surface brightness extended structures.   
\par
We note that the redshift distribution of the 5030 spiral galaxies has median $z_{\rm median} = 0.07$ with $>90\%$ and $>99\%$ 
sources at $z \geq$ 0.032 and $z \geq$ 0.02 respectively. The search radius of 600$^{{\prime}{\prime}}$ corresponds 
to 807\,kpc, 386\,kpc and 245\,kpc at $z_{\rm median} = 0.07$,  $z_{\rm 90{\%}} = 0.032$ and  $z_{\rm 99{\%}} = 0.02$, 
respectively. 
This implies that our search radius of 600$^{{\prime}{\prime}}$ will pick up emitting components separated by up to 490\,kpc 
and 772\,kpc in $> 99\%$ and $> 90 \%$ of sources, respectively. 
Therefore, even at low-redshifts, our search radius will effectively pick up giant radio galaxies in which the centroids of 
emitting components are separated by a few hundreds of kpc, and where the total end-to-end linear projected size is $\sim 1$\,Mpc 
\citep{Schoenmakers01,Saripalli05}. 
We also note that our search radius is larger than that used in previous studies ({\eg}450$^{{\prime}{\prime}}$; \cite{Best05}) 
to search for radio galaxies associated with SDSS galaxies. \par
Our multistage search process involving step (i) and (ii) yields a total of six double-lobe 
radio galaxies associated with spiral galaxies.
One of these six sources J1649+2635 has already been reported by \cite{Mao15}. 
To further check the spiral nature of the host galaxies, we use the `Galaxy Zoo' catalogue 
and consider only galaxies with a probability of being spiral 
$\geq$ 0.8. 
We note that two of our six sources {\ie}J1008+5026 (P$_{\rm spiral} = 0.33$), 
J1450-0106 (P$_{\rm spiral} = 0.66$), appear to be bulge-dominated spheroidals with 
low probability of being them spiral. 
These sources are classified as spirals in the \cite{Meert15} catalogue mainly due to the presence of a 
weak exponential component which permits a statistically better fit to the optical surface-brightness profile.  
Therefore, we do not consider these two galaxies as confirmed spirals and exclude them 
from further study. 
Thus, our final sample of double-lobe radio galaxies hosted in spiral galaxies consists of 
four sources {\ie}J0836+0532, J1159+5820, J1352+3126 and J1649+2635 (see Table 1 and Figure 1).
\par
We note that our search for spiral-host radio galaxies is not complete, as we miss 
core-less double-lobe sources that are either lobe-dominated or relic radio galaxies with a weak (or absent) undetected core component. 
An attempt to search for core-less double-lobe radio sources around optical sources 
gives random matches without confirmed positional coincidence.
We also miss those radio galaxies in which lobes are 
resolved out in FIRST and whole source remain unresolved in the NVSS 
({\ie}total end-to-end projected size is less than the NVSS beam-size $\sim$ 45$\arcsec$). 
Furthermore, our sample of spiral galaxies is based on the morphological classification given in \cite{Meert15} and 
therefore, we miss those spirals that are not listed in 
this catalogue. For example, J1409-0302 (Speca; \cite{Hota11}) is reported as a spiral-hosted radio galaxy 
but it has been assigned `uncertain' morphology in the \cite{Meert15} catalogue. 
We also note that, 0313-192 \citep{Ledlow98} as well as J2345-0449 \citep{Bagchi14} are not included in \cite{Meert15} 
since they fall outside the SDSS DR7 footprint. 
\par
In order to check that our sources are not simple chance matches, we estimate the probability of chance coincidence for 
a double-lobe radio galaxy from FIRST and NVSS to match with a spiral galaxy within a search radius of 3$\arcsec$ centred at the optical position. 
Surface source densities in the FIRST and NVSS are $\sim$ 90 deg$^{-2}$ \citep{White97} and 50 deg$^{-2}$ \citep{Condon98}, respectively.  
We assume that $\sim$70$\%$ of FIRST detected radio sources at the faintest level are AGN, 
and only 10 $\%$ of these are radio$-$loud \citep{Ivezic02}. 
Although, at higher flux densities, the fraction of AGN increases and the source density decreases.   
Furthermore, nearly two-thirds ($\sim$ 67$\%$) of all the radio-loud AGN can be recognized as double-lobe radio galaxies 
owing to the projection effect 
({\ie}in general FR-I, FR-II radio galaxies have typical viewing angle of jet $\gtrsim$ 30$^{\circ}$ \citep{Ghisellini93} and 
their counterparts seen at smaller viewing angles are likely to be classified as BL Lacs and radio-loud quasars 
as per the AGN unification model \citep[]{Urry95}).  
Therefore, only $\leq$ 4.7$\%$ FIRST detected radio sources with average surface density 
($\leq$ 90 $\times$ 4.7$\%$) $\leq$ 4.23 deg$^{-2}$ 
are expected to be double-lobe radio galaxies. NVSS will have an even lower double-lobe radio galaxy source density 
due to its overall lower surface source density. 
Considering a circle of radius 3$\arcsec$ around the optical position, the probability of a double-lobe radio galaxy falling within it by chance 
is $\pi$(3.0/3600)$^{2}$ $\times$ 4.23. And, since there are a total of 187005 spiral galaxies, the cumulative number of chance matches  
is $\pi$(3.0/3600)$^{2}$ $\times$ 4.23 $\times$ 187005 $\simeq$ 1.7.     
The total number of six double-lobe radio galaxies showing a FIRST central component within 3$\arcsec$ 
in our sample is higher than the number of chance matches. 
However, we note that one to two chance matches are not in fact unexpected. 
We further note that our spiral-host double-lobe radio galaxies are optical AGN 
(see Table~\ref{table:LineRatio} and Figure~\ref{fig:KewleyPlot}). 
Therefore, probability of double-lobe radio galaxies positionally matching to optical AGN by chance would be even lower, 
because only a fraction of spiral galaxies in the Meert catalog are AGN.
\begin{figure*}
\centering
\begin{subfigure}[b]
{0.3\textwidth}
\includegraphics[angle=0,width=7.5cm,trim={0.0cm 0.0cm 0.0cm 0.0cm},clip]{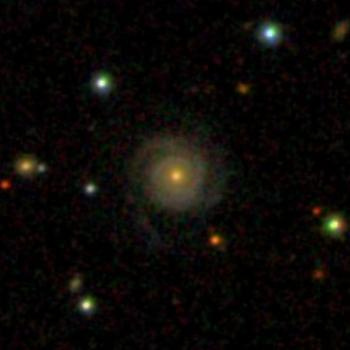}
\caption{J0836+0532}
\label{fig:J0836+0532}
\end{subfigure}
\hspace{2.5cm}
{
\begin{subfigure}[b]
{0.3\textwidth}
\includegraphics[angle=0,width=7.5cm,trim={0.0cm 0.0cm 0.0cm 0.0cm},clip]{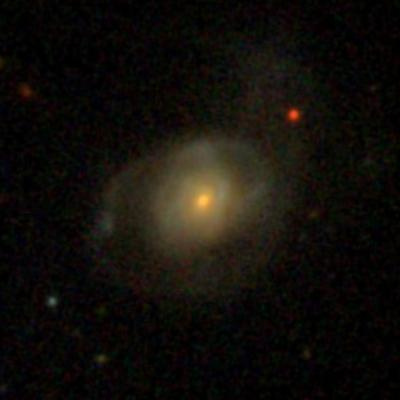}
\caption{J1159+5820}
\label{fig:J1159+5820}
\end{subfigure}}
\begin{subfigure}[b]
{0.3\textwidth}
\includegraphics[angle=0,width=7.5cm,trim={0.0cm 0.0cm 0.0cm 0.0cm},clip]{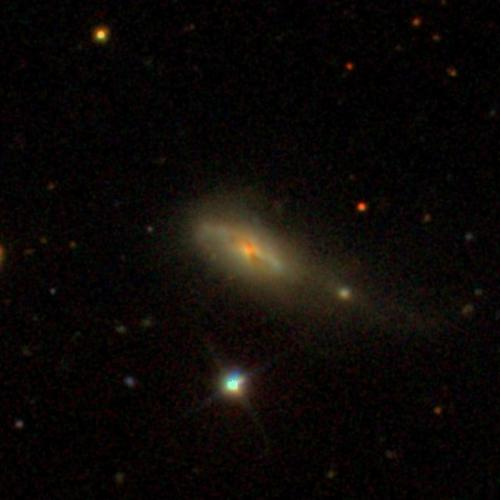}
\caption{J1352+3126}
\label{fig:J1352+3126}
\end{subfigure}
\hspace{2.5cm}
{
\begin{subfigure}[b]
{0.3\textwidth}
\includegraphics[angle=0,width=7.5cm,trim={0.0cm 0.0cm 0.0cm 0.0cm},clip]{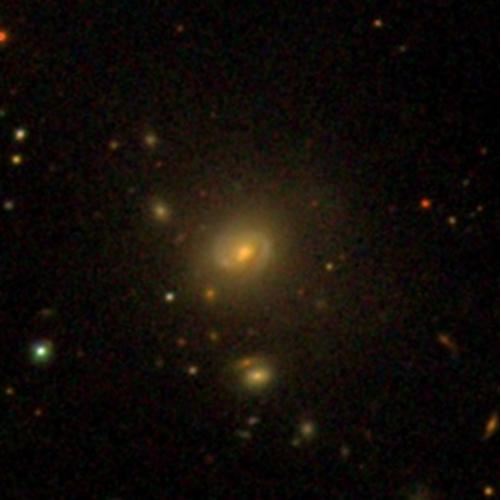}
\caption{J1649+2635}
\label{fig:J1649+2635}
\end{subfigure}}
\caption{SDSS optical images}
\label{fig:SDSSImages} 
\end{figure*}

\section{Properties of host galaxies}
\subsection{Optical surface brightness profile}
In order to check the morphological parameters of the host galaxies, we use the best fit 
S\'ersic model parameters derived from $r$-band SDSS images, as given in the \cite{Meert15} catalogue (see Table~\ref{table:OptMorph}).
The optical surface brightness profile of a spiral galaxy having bulge and disk components was fitted with a 
two-component model, {\ie}a S\'ersic model for the central bulge component, and a disk component fitted with an 
exponential law defined as $I(r) = I_{\rm d} \exp(-r/R_{\rm d})$, where $R_{\rm d}$ is the scale radius, and $I_{\rm d}$ is the central surface brightness. 
The S\'ersic model is defined as $I(r) = I_{e} \exp(-b_{n} [(r/R_{e})^{1/n} - 1])$ \citep{Sersic63}; 
where $b_{\rm n}$ = 1.9992$n$ - 0.3271 and is valid for $0.5 < n < 10$ (using the approximation from \cite{Capaccioli89}).
For $n = 4$, the S\'ersic model reduces to the de Vacouleurs model \citep{DeVaucouleurs48}. 
The parameters that define the S\'ersic profile are S\'ersic index ($n$), half-light radius ($R_{\rm e}$), and surface brightness 
($I_{\rm e}$) at $R_{\rm e}$. 
The SDSS r-band optical surface brightness profiles of our sample sources (see Figure~\ref{fig:J0836+0532} to \ref{fig:J1649+2635}), 
best fitted with the S\'ersic model plus an exponential disk component, are
taken from \cite{Meert15}\footnote{http://shalaowai.physics.upenn.edu/$\sim$ameert/fit{\_}catalog/view{\_}galaxies/}.

\subsection{Optical spectral properties} 
Optical spectral properties can be used to reveal the nature of the host galaxy. 
We examine the SDSS spectra of our sample sources. 
These are extracted from fibers of 3$\arcsec$ diameter positioned at the centres of galaxies, and are therefore 
dominated by the light from the central region of galaxies. 
All our sample sources are at $z > 0.04$ and thus capture $> 20\%$ of the galaxy light \citep[see][]{Kewley05}. 
Thus, SDSS spectra sample the light from the nucleus (AGN) as well as from the central part of the galaxy. 
Spiral galaxies are expected to show strong blue continuum due to the presence of young stellar populations.  
We also use emission line ratio diagnostics proposed by \cite{Kewley06} to identify if any emission lines present are due to
star formation or AGN. 
Table~\ref{table:LineRatio} lists the emission line ratios 
({\ie}[O III]/H$_{\beta}$, [N II]/H$_{\alpha}$, [S II]/H$_{\alpha}$ and [O I]/H$_{\alpha}$) 
and the spectral classification based on the emission line diagnostics \citep{Baldwin81,Kewley06}).
The SDSS optical spectra of our sample sources, presented in 
figure~\ref{fig:J0836+0532} to \ref{fig:J1649+2635}, are taken from the SDSS website\footnote{http://www.sdss.org/dr12/spectro/}.

\subsection{Star formation rate}
Star Formation Rate (SFR) is one of the characteristic properties of galaxies.
We estimate SFRs for our sample sources using mid-IR and UV continuum, whenever available. 
It has been shown that the monochromatic mid-IR luminosity is tightly correlated with total 
IR luminosity and can be used to estimate SFR \citep[{\eg}][]{Rieke09}. 
All of our sample sources are detected in the mid-IR by the Wide-field Infrared Survey Explorer (WISE; \cite{Wright10}), and we use 
the 22 $\mu$m (W4 band) luminosity to estimate SFRs. 
The WISE colours of our sample sources are [3.4] - [4.6] $<$ 0.8 (Vega magnitudes), indicating that the mid-IR emission is dominated by star 
formation and AGN contamination is not significant \citep[see][]{Stern12}.  
To obtain mid-IR flux densities from magnitudes, we first converted WISE Vega magnitudes into AB magnitudes 
using conversion factors (m$_{\rm AB}$ = m$_{\rm Vega}$ + $\Delta$m; 
where ${\Delta}$m is 2.683, 3.319, 5.242 and 6.604 for W1, W2, W3 and W4 bands, respectively) 
given in the WISE data release document\footnote{http://wise2.ipac.caltech.edu/docs/release/prelim/expsup/sec4{\_}3g.html}. 
We estimate total IR (8 - 1000 $\mu$m) luminosity (L$_{\rm IR}$) from the WISE 22 $\mu$m luminosity using the 
full range of templates in the libraries of \cite{Chary01} and \cite{Dale02}. 
The total IR luminosities were then converted into SFRs using the \cite{Kennicutt98} law 
{\ie}SFR (M$_{\odot}$ yr$^{-1}$) = 4.5 ${\times}$ 10$^{-44}$ $L_{\rm 8 - 1000~{\mu}m}$ (ergs s$^{-1}$). 
We adopt the median SFR value from the full template set as the SFR of galaxy, and the error bars are 
upper/lower limits considering the full template set. 
Table~\ref{table:MIR} lists the mid-IR WISE magnitudes, colours, IR luminosities and SFRs of our sample sources. \par 
Two of our sample sources are detected both in the Near-UV (NUV; 1750-2750~{\AA}) and Far-UV (FUV; 1350-1750~{\AA}) bands of 
GALEX \citep{Martin05} as the part of All-sky Imaging Survey (AIS). 
J1649+2635 is detected only in the NUV band. 
Table~\ref{table:Mags} lists UV magnitudes for our sample sources. 
Since UV emission in star forming galaxies is dominated by young massive stars, SFR scales with increasing UV luminosity. 
We use UV luminosities to estimate SFRs using \cite{Kennicutt98} equation : 
SFR (M$_{\odot}$ yr$^{-1}$) = 1.4 ${\times}$ 10$^{-28}$ $L_{\rm UV}$ (ergs s$^{-1}$). 
We use observed UV luminosities to estimate SFRs, and therefore these SFRs are only lower limits. 
The extinction corrections are problematic, as the spatial distribution of dust causing extinction can be patchy and UV emission 
may be primarily from regions with less obscuration \citep{Calzetti94}.  
We also caution that UV emission may have some contribution from processes 
such as scattered light from the AGN, in addition to that from star formation \citep[{\eg}][]{Tadhunter02}. 
In Table~\ref{table:MIR}, we list the SFRs derived from IR and UV observed luminosities. 
Notably, SFRs derived from UV luminosities for three sources are systematically lower than those from IR luminosities.
\subsection{Supermassive black hole mass}
The mass of the supermassive black hole is one of the fundamental parameters of an AGN host galaxy system.  
We estimate black hole masses for our sample sources using the black hole mass--bulge luminosity relation \citep{Kormendy95,Magorrian98}. 
Absolute bulge magnitudes of our sample sources are taken from \cite{Simard11} who present bulge-disk decomposition for SDSS DR7 galaxies. 
The absolute bulge magnitudes are converted into luminosities, which are in turn used to estimate black hole masses. 
We use the empirical black hole mass--bulge luminosity relations given in \cite{McConnell13} and \cite{Bentz09}.
\cite{McConnell13} present scaling relations for black hole mass and host galaxy properties for early and late type galaxies. 
As a sanity check, we also use the empirical black hole mass--bulge luminosity relation given in \cite{Bentz09} for a sample of AGN. 
\cite{Bentz09} use black hole mass measurements based on reverberation mapping and bulge luminosities from two-dimensional 
decompositions of Hubble Space Telescope (HST) images. 
Table~\ref{table:Msmbh} lists bulge magnitudes, luminosities and black hole mass estimates for our sample sources.   
Black hole mass estimates derived from the scaling relations for late type galaxies and AGN are likely to be better estimates for 
our sample sources as the hosts of our sources are spirals with AGN. 

\subsection{Environment}
It is known that the nature of a galaxy can be influenced by the large scale surrounding environment.
To check if our sample sources are residing in groups or clusters of galaxies, 
we look at the distribution of galaxies around our sample sources. 
Noting that the typical velocity dispersion in a cluster of galaxies is $<$ 1000 km s$^{-1}$ \citep{Becker07}, 
we consider galaxies only within a redshift range of $\Delta$cz = $\pm$1000 km s$^{-1}$ ({\ie}$\Delta$z = $\pm$0.0033) {\it w.r.t.} the 
redshift (z) of our sample galaxies.
We use only galaxies and QSOs with redshifts given in the SDSS DR10 spectroscopic catalogue. 
The SDSS catalogue includes photometric redshift estimates for more galaxies, including fainter ones, but the redshift errors are much larger than 
$\Delta$cz = $\pm$1000 km s$^{-1}$, and this makes them unsuitable for our analysis. 
Clusters and groups of galaxies typically have sizes up to few Mpc, but since we do not know the position 
of our source {\it w.r.t.} any parent cluster, we investigate the distribution of galaxies in 
a large circular region of radius varying from 0.5 Mpc to 4.0 Mpc. 
Galaxy groups typically contain a few galaxies to a few tens of galaxies in a diameter as large as 1.0 to 2.0 Mpc, while 
systems with more galaxies are defined as clusters \citep{Lin04,Tago08}. 
Although, we caution that there is no sharp dividing line between a large group and a small cluster.  
Table~\ref{table:Environ} lists the number of galaxies with redshifts z${\pm}{\Delta}$z that fall within the projected circular regions of 
the radii of 0.5 Mpc, 1.0 Mpc, 2.0 Mpc and 4.0 Mpc centred at our sample galaxies. 
Furthermore, we also use the \cite{Tempel14} catalogue of groups and clusters to examine the association of our sample sources with over-densities. 
Using SDSS DR10 data \cite{Tempel14} identified galaxy groups and clusters based on a modified friends-of-friends method and present a 
flux (m$_{\rm r}$ = 17.77) and volume limited catalogue.
\cite{Tempel14} considered galaxy group as a system of $\geq$ 3 galaxies where galaxies are linked into the system 
using a linking length ($\sim$ 0.25 Mpc/h in the projected distance at z = 0) that varies with the distance.
The group finding algorithm is explained in \citep{Tago08,Tago10}.
We list estimated positions, sizes, masses and densities of the groups associated with our sample galaxies in Table~\ref{table:Environ}.
The presence of a cluster can also be found by the detection of extended X-ray emitting hot intracluster gas. 
Therefore, we checked archival {\em RoSAT} All Sky Survey (RASS; \cite{Voges99,Voges2000}) 
to examine the presence of soft X-ray (0.1 - 2.4 keV) emission around our sample sources.    
We find that none of our sample sources are detected in the RASS. 

\begin{table*}
\begin{minipage}{140mm}
\caption{Our sample sources and their optical morphological parameters}
\begin{tabular}{@{}cccccccccccc@{}}
\hline
Source       &  RA         &  Dec        & Redshift  & Best fit  & ${\chi}^{2}$/dof &  S\'ersic     & r$_{\rm bulge}$ & r$_{\rm disk}$ & B/T & b/a  & P$_{\rm spiral}$ \\ 
name         & (h m s)     &  (d m s)    &           &  model    &        &  index (n)            &     ({\arcsec}) &   ({\arcsec})   &      &      &                       \\ \hline
J0836+0532   & 08 36 55.9  & +05 32 42.0 & 0.099     & Ser + Exp &  1.03  &  0.83$\pm$0.06        &  0.64$\pm$0.01  &  3.82$\pm$0.04  & 0.15 & 0.91 &  0.95  \\
J1159+5820   & 11 59 05.8  & +58 20 35.5 & 0.054     & Ser + Exp &  1.07  &  7.69$\pm$0.20        &  29.42$\pm$2.06 & 5.07$\pm$0.03   & 0.75 & 0.75 &  0.81   \\
J1352+3126   & 13 52 17.8  & +31 26 46.3 & 0.045     & Ser + Exp &  1.18  &  5.83$\pm$0.27        &  51.34$\pm$6.22 & 7.78$\pm$0.02   & 0.61 & 0.60 &  0.86   \\
J1649+2635   & 16 49 23.9  & +26 35 02.7 & 0.055     & Ser + Exp &  1.16  &  2.81$\pm$0.03        &  7.57$\pm$0.12  & 20.82$\pm$0.76  & 0.63 & 0.71 &  0.97   \\ \hline
\end{tabular}
\label{table:OptMorph} 
\\
Note - Best fit model parameters are based on \cite{Meert15} catalogue. Ser : S\'ersic model; Exp : Exponential model. 
r$_{\rm bulge}$ is the bulge radius, r$_{\rm disk}$ is the disk radius, B/T is the bulge-to-total light ratio from fit, 
b/a is the axis ratio (semi-minor/semi-major) of the total fit, 
P$_{\rm spiral}$ is the debiased probability for host galaxy to be a spiral according to the Galaxy Zoo catalogue.  
\end{minipage}
\end{table*}

\begin{table*}
\begin{minipage}{140mm}
\caption{UV and optical magnitudes}
\begin{tabular}{@{}ccccccccc@{}}
\hline
Survey       & \multicolumn{2}{c} {GALEX}  & \multicolumn{5}{c} {SDSS}  &   \\  
Band         &    FUV   & NUV              & u  &  g  &  r  &  i  &  z    &  u - r \\ \hline
{\it Source} &                &                 &       &       &       &       &        &    \\
J0836+0532   &                &                 & 18.187$\pm$0.034 & 16.432$\pm$0.004 & 15.545$\pm$0.004 & 15.109$\pm$0.004 & 14.756$\pm$0.010 &  2.64  \\
J1159+5820   & 19.73$\pm$0.17 & 18.76$\pm$0.04  & 16.670$\pm$0.015 & 14.775$\pm$0.002 & 13.982$\pm$0.002 & 13.553$\pm$0.002 & 13.220$\pm$0.003 &  2.69  \\
J1352+3126   & 20.49$\pm$0.27 & 18.65$\pm$0.05  & 16.857$\pm$0.017 & 15.056$\pm$0.003 & 14.125$\pm$0.002 & 13.615$\pm$0.002 & 13.261$\pm$0.004  & 2.74  \\
J1649+2635   &                & 20.63$\pm$0.15  & 17.653$\pm$0.027 & 15.496$\pm$0.003 & 14.478$\pm$0.002 & 13.973$\pm$0.002 & 13.581$\pm$0.004 &  3.17  \\ \hline
\end{tabular}
\label{table:Mags} 
\\
Note - GALEX magnitudes are from All-sky Imaging Survey (AIS)\footnote{http://galex.stsci.edu/GalexView/}. 
SDSS magnitudes are model based.
\cite{Strateva01} proposed that optical colour (u - r) can effectively separate red (u - r $\leq$ 2.22) and blue galaxies (u - r $>$ 2.22) 
in the SDSS data.    
\end{minipage}
\end{table*}
\begin{table*}
\begin{minipage}{140mm}
\caption{WISE mid-IR magnitudes, colours, luminosities and SFRs}
\begin{tabular}{@{}ccccccccc@{}}
\hline
Source       &  W1        &  W2   &   W3  &  W4  &  W1 - W2 &   L$_{\rm IR}$ & SFR$_{\rm IR}$ & SFR$_{\rm UV}$ \\
name         & (3.4 $\mu$m) & (4.6 $\mu$m) & (12 $\mu$m) & (22 $\mu$m) &   & (L$_{\odot}$) & (M$_{\odot}$ yr$^{-1}$) & (M$_{\odot}$ yr$^{-1}$) \\ \hline
J0836+0532   & 12.87 & 12.42 & 9.61  & 7.49 &  0.45  &   5.80$^{+5.61}_{-2.18}$ $\times$ 10$^{10}$ & 9.99$^{+9.66}_{-3.76}$ &       \\
J1159+5820   & 11.82 & 11.75 & 9.38  & 7.38 &  0.07  &   1.68$^{+1.62}_{-0.67}$ $\times$ 10$^{10}$ & 2.89$^{+2.78}_{-1.16}$ &  1.50  \\
J1352+3126   & 11.33 & 10.79 & 8.12  & 5.96 &  0.54  &   4.31$^{+4.13}_{-1.74}$ $\times$ 10$^{10}$ & 7.41$^{+7.12}_{-2.99}$ &  0.97  \\
J1649+2635   & 11.79 & 11.80 & 9.69  & 8.01 & -0.01  &   9.68$^{+9.31}_{-3.87}$ $\times$ 10$^{9}$  & 1.67$^{+1.60}_{-0.67}$ &  0.20  \\ \hline
\end{tabular}
\label{table:MIR} 
\\
Note - WISE magnitudes are in the Vega system. L$_{\rm IR}$ is the 8 - 1000 $\mu$m luminosity estimated from 22 $\mu$m luminosity using the 
full range of templates in the libraries of \cite{Chary01} and \cite{Dale02}. SFRs are estimated using the \cite{Kennicutt98} law. 
\end{minipage}
\end{table*}
\begin{table*}
\centering
\begin{minipage}{160mm}
\caption{Optical emission line fluxes and ratios from the SDSS}
\resizebox{18cm}{!}{
\begin{tabular}{@{}cccccccccccc@{}}
\hline
Source &  H$_{\beta}$ & [O III]    & [O I]      & H$_{\alpha}$ & [N II]      & [S II]      & log([O III]/H$_{\beta}$) & log([N II]/H$_{\alpha}$) & log([S II]/H$_{\alpha}$) & log([O I]/H$_{\alpha}$) & BPT  \\
name   & $\lambda$ 4863 {\AA} & $\lambda$ 5007 {\AA} & $\lambda$ 6300 {\AA} & $\lambda$ 6563 {\AA} & $\lambda$ 6584 {\AA} & $\lambda$ 6717 {\AA}  &        &              &           &          & class \\ \hline
J0836+0532   &  26.66 &   329.77   &  45.89     &  163.32      &  250.56    &  65.44      &       1.09               &     0.19                 &   -0.39                  &   -0.55             & Seyfert~/~HERG \\
J1159+5820   &  13.69 &   167.01   &  70.17     &   234.67     &  396.72    & 157.46      &       1.09               &     0.23                 &   -0.17                  &   -0.52             & Seyfert~/~HERG \\
J1352+3126   &  95.25 &  178.13    &  202.62    &  982.84      & 1022.29    & 622.74      &       0.27               &     0.02                 &   -0.20                  &   -0.69             & LINER~/~LERG  \\
J1649+2635   &        &  50.84     &  38.51     &  128.94      &  145.25    &  56.00      &                          &     0.05                 &   -0.36                  &   -0.52             &              \\ \hline
\end{tabular}
}
\label{table:LineRatio} 
\\
Note - Fluxes are in units of 10$^{-17}$ erg cm$^{-2}$ s$^{-1}$. 
Classification is based emission line ratio diagnostics 
(famously known as `BPT' diagnostics) proposed by \cite{Baldwin81} \citep[see][]{Kewley06}. 
Radio galaxies can also have emission line ratios similar to low luminosity AGN ({\ie}Seyfert and LINER galaxies), but 
High Excitation Radio Galaxies (HERGs) and Low Excitation Radio Galaxies (LERGs) occupy different regions in the BPT diagrams 
\citep{Buttiglione10}.
\end{minipage}
\end{table*}

\begin{table*}
\begin{minipage}{140mm}
\caption{Bulge magnitudes, luminosities and black hole mass estimates}
\begin{tabular}{@{}cccccccc@{}}
\hline
Source & Mag$_{\rm g{\_}bulge}$  &  Mag$_{\rm r{\_}bulge}$  &   Mag$_{\rm v{\_}bulge}$  &  L$_{\rm v{\_}bulge}$ &  M$_{\rm SMBH}^{\rm a}$ (M$_{\odot}$)   &  M$_{\rm SMBH}^{\rm b}$ (M$_{\odot}$)   & M$_{\rm SMBH}^{\rm c}$ (M$_{\odot}$)\\
name   &                      &                       &                        &    (L$_{\odot}$)   & $\alpha$ = 9.23, $\beta$ = 1.11 & $\alpha$ = 9.10, $\beta$ = 0.98 & $\alpha$ = 7.98, $\beta$ = 0.80 \\ \hline
J0836+0532 & -20.98$\pm$0.03 & -21.95$\pm$0.02 & -21.55$\pm$0.02 & 3.63$\pm$0.07 $\times$ 10$^{10}$ &  5.5$\pm$1.7 $\times$ 10$^{8}$ & 4.7$\pm$2.7 $\times$ 10$^{8}$ & 2.7$\pm$1.0 $\times$ 10$^{8}$ \\
J1159+5820 & -20.96$\pm$0.01 & -22.02$\pm$0.01 & -21.58$\pm$0.01 & 3.73$\pm$0.03 $\times$ 10$^{10}$ &  5.7$\pm$1.3 $\times$ 10$^{8}$ & 4.8$\pm$2.7 $\times$ 10$^{8}$ & 2.8$\pm$1.0 $\times$ 10$^{8}$ \\
J1352+3126 & -19.66$\pm$3.26 & -21.13$\pm$0.18 & -20.52$\pm$1.37 & 1.41$\pm$1.78 $\times$ 10$^{10}$ &  1.9$\pm$0.4 $\times$ 10$^{8}$ & 1.9$\pm$1.1 $\times$ 10$^{8}$ & 1.7$\pm$0.5 $\times$ 10$^{8}$ \\
J1649+2635 & -20.47$\pm$0.01 & -21.43$\pm$0.01 & -21.04$\pm$0.01 & 2.27$\pm$0.02 $\times$ 10$^{10}$ &  3.3$\pm$0.7 $\times$ 10$^{8}$ & 2.9$\pm$1.7 $\times$ 10$^{8}$ & 1.8$\pm$0.7 $\times$ 10$^{8}$ \\ \hline
\end{tabular}
\label{table:Msmbh} 
\\
Note - g-band and r-band bulge magnitudes (Mag$_{\rm g{\_}bulge}$ and Mag$_{\rm r{\_}bulge}$) are taken from \cite{Simard11} catalogue, where 
bulge is fitted with S\'ersic model with S\'ersic index (n) as a free parameter.
V-band bulge magnitude is estimated from g-band and r-band magnitudes using \cite{Jester05} magnitude 
conversion equation (V = g - 0.58*(g-r) - 0.01). 
Black hole masses (M$_{\rm SMBH}$) are estimated using black hole mass - bulge luminosity relation 
(log$\left(\frac{\rm M_{\rm SMBH}}{\rm M{_{\odot}}}\right)$ = $\alpha$ + $\beta$ log$\left(\frac{\rm Lv{\_}bulge}{\rm 10^{11}~L{_{\odot}}}\right)$; 
where $\alpha$ = 9.23$\pm$0.10, $\beta$ = 1.11$\pm$0.13 for early type galaxies 
and $\alpha$ = 9.10$\pm$0.23, $\beta$ = 0.98$\pm$0.20 for late type galaxies with 
L$_{\rm v{\_}bulge}$ $\leq$ 10$^{10.8}$) given in \cite{McConnell13} and 
(log$\left(\frac{\rm M_{\rm SMBH}}{\rm M{_{\odot}}}\right)$ = $\alpha$ + $\beta$ log$\left(\frac{\rm Lv{\_}bulge}{\rm 10^{10}~L{_{\odot}}}\right)$; 
$\alpha$ = 7.98$\pm$0.06, $\beta$ = 0.80$\pm$0.09 derived for AGN) given in \cite{Bentz09}. 
$^{\rm a}$ : using \cite{McConnell13} relation for early type galaxies; $^{\rm b}$ : using \cite{McConnell13} relation for late type galaxies; 
and $^{\rm c}$ : using \cite{Bentz09} relation for AGN.         
\end{minipage}
\end{table*}

\begin{table*}
\begin{minipage}{140mm}
\caption{Radio flux densities}
\begin{tabular}{@{}cccccc@{}}
\hline
Survey       &   GB6$^{1}$      & FIRST & NVSS & B2$^{2}$ & VLSS$^{2}$ \\  
Band         &   4.85 GHz & 1.4 GHz & 1.4 GHz & 408 MHz & 74 MHz \\  
             &   (mJy)    & (mJy)   & (mJy)   & (mJy)   & (mJy)  \\ \hline
{\it Source} &       &       &       &        &         \\
J0836+0532   &       & 23.81$\pm$0.36 & 62.4$\pm$2.3  &        &         \\
J1159+5820   &  257.0$\pm$38.6  & 8.61$\pm$0.56 &  338.1$\pm$7.5 &        &         \\
J1352+3126   &  1940.0$\pm$291 &  4119.97$\pm$5.73  &  4844.2$\pm$145.8 &  10460$\pm$850 &         \\
J1649+2635   &  63$\pm$10 &  101.97$\pm$0.84  & 155.4$\pm$5.4 & 331$\pm$7 & 610$\pm$100     \\ \hline
\end{tabular}
\label{table:Radioflux} \\
Notes - Only total integrated radio flux densities are listed. $^{1}$ : Green Bank survey \citep{Gregory91}; $^{2}$ : Bologna survey \citep{Colla72}; 
$^{3}$ : VLA Low-Frequency Sky Survey \citep{Cohen07b}.
\end{minipage}
\end{table*}
%
%
\begin{table*}
\centering
\begin{minipage}{140mm}
\caption{Our sample sources and previous examples of spiral host radio galaxies}
\begin{tabular}{@{}cccccccccc@{}}
\hline
Source       &  RA         &  Dec        & Redshift  & scale           & size   & L$_{\rm 1.4~GHz}$    & FR       & Environment & Reference  \\ 
name         &             &             &           &(kpc/${\arcsec}$)& (kpc)  & (W Hz$^{-1}$)        & type     &             &       \\ \hline
J0836+0532   & 08 36 55.9  & +05 32 42.0 & 0.099     &  1.756          &  420   & 1.53 $\times$ 10$^{24}$ & FR-II  &   Field          & 1  \\ 
J1159+5820   & 11 59 05.8  & +58 20 35.5 & 0.054     &  1.037          &  392   & 2.26 $\times$ 10$^{24}$ & FR-II &  Field         & 1  \\ 
J1352+3126   & 13 52 17.8  & +31 26 46.3 & 0.045     &  0.877          &  335   & 2.26 $\times$ 10$^{25}$ & FR-II  &  Field        & 1  \\ 
J1649+2635   & 16 49 23.9  & +26 35 02.7 & 0.055     &  1.046          &  86   & 1.07 $\times$ 10$^{24}$ & FR-II  &  group/BCG    & 1, 4  \\    
J0315-1906 (0313-192) & 03 15 52.1 & -19 06 44.0 & 0.067  &  1.268 & 200   & 1.0 $\times$ 10$^{24}$ & FR-I  &  Abell 428  & 2  \\ 
J1409-0302 (Speca)    & 14 09 48.8 & -03 02 32.5 & 0.138  &  2.411 & 1000 & 7.0 $\times$ 10$^{24}$ & FR-II &  BCG        & 3  \\ 
J2345-0449   & 23 45 32.7  & -04 49 25.3 & 0.076     &  1.423      & 1600 & 2.5 $\times$ 10$^{24}$ & FR-II &  Field/CL outskirt & 5 \\ \hline
\end{tabular}
\label{table:Sample} 
Note : Sizes are  end-to-end linear projected distances in FIRST or NVSS images whichever show clear double-lobe radio morphology. 
Total 1.4 GHz radio luminosities are from NVSS as extended emission is often missed in FIRST. \\ 
Reference, 1 : This work; 2 : \cite{Ledlow98}; 3 : \cite{Hota11}; 4 : \cite{Mao15}; 5 : \cite{Bagchi14}.
\end{minipage}
\end{table*}
\begin{table*}
\centering
\begin{minipage}{180mm}
\caption{FIRST and NVSS radio component parameters}
\resizebox{18cm}{!}{
\begin{tabular}{@{}cccccccccccccc@{}}
\hline
               & \multicolumn{7}{c} {FIRST}   & \multicolumn{6}{c} {NVSS}   \\  \hline
               &  S$_{\rm p}$ & S$_{\rm int}$ &  Major & Minor & PA & R$_{\rm off}$ & Morphology & S$_{\rm int}$ &  Major & Minor & PA & R$_{\rm off}$ & Morphology \\
               &   (mJy/b)    &   (mJy)       & ({\arcsec})& ({\arcsec})& (deg) & ({\arcsec})&           &  (mJy)        &({\arcsec})&({\arcsec})& (deg) &({\arcsec})&             \\ \hline
J0836+0532     &              &               &        &       &    &               &  C + 2L    &               &        &       &    &               &  C +2L       \\
 Core          &   9.32       &  16.45        &  7.55  &  1.26 & 97.2 & 1.1  &            &  23.2 &  26.0 & 24.6 &  73.5 & 1.04 &             \\
 lobe 1 (SW)   &   2.19       &  7.36         &  11.15 &  5.78 & 13.3 & 80.2 &            &  25.4 &  35.2 & 27.0 & -19.2 & 74.5 &             \\
 lobe 2 (NE)   &              &               &        &       &       &      &           &  13.8 &  34.3 & $<$ 37.0& -52.1 & 80.7 &          \\
               &              &               &        &       &       &      &           &       &       &      &       &        &         \\
J1159+5820     &              &               &        &       &       &      &  C + 2L   &       &       &      &       &        & C + 2L  \\
  Core         &   2.06       &  2.11         &  2.21  &  0.00 & 113.6 & 1.1  &           &       &       &      &       &        &         \\
lobe 1 (new)   &   1.95       &  2.14         &  3.56  &  0.00 & 120.2 & 10.0 &           &       &       &      &       &        &         \\
lobe 2 (new)   &   2.92       &  4.36         &  4.77  &  2.76 & 42.1  & 12.5 &           &       &       &      &       &        &          \\
lobe 1 (old)C1 &              &               &        &       &       &      &           & 145.4 & 122.0 & 57.4 & -11.9 & 118    &         \\
lobe 1 (old)C2 &              &               &        &       &       &      &           & 10.6  & 87.0  &$<$29.8& -42.0& 115    &         \\
lobe 1 (old)C3 &              &               &        &       &       &      &           & 66.3  & 127.1 & 16.7 & 78.1  &  62.6  &         \\
lobe 2 (old)C1 &              &               &        &       &       &      &           & 36.0  & 126.4 & 65.0 & -31.4 & 159    &         \\
lobe 2 (old)C2 &              &               &        &       &       &      &           & 76.1  & 112.6 & 29.3 & 80.9  & 116    &         \\
lobe 2 (old)C3 &              &               &        &       &       &      &           & 3.7   &$<$43.6&$<$41.6&      & 35.8   &          \\
               &              &               &        &       &       &      &           &       &       &       &      &        &          \\
J1352+5348     &              &               &        &       &       &      &  C + 2L   &       &       &       &      &        &  C + 2L     \\
 Core          &  3334.28     &   3709.13     & 2.50   & 0.75  & 100.3 & 0.6  &           &4085.4 & 14.8  &$<$17.9& -15.5& 0.44   &             \\
 lobe 1 (NW)   &  33.31       &    393.3      & 14.58  & 5.28  & 14.6  & 85.0 &           & 629.1 & 42.9  & 20.8  & -30.6& 76.5   &             \\
 lobe 2 (SE)   &  6.74        &   17.54       & 6.95   & 0.00  & 148.4 & 97.5 &           & 129.7 & 83.2  & 30.7  & -23.3& 107    &             \\
               &              &               &        &       &       &      &           &       &       &       &      &        &             \\
J1649+2635     &              &               &        &       &       &      &  C + 2L   &       &       &       &      &        &  PS        \\
 Core          &  6.34        &  8.59         &  4.64  &  1.29 &  47.1 & 1.4  &           & 155.4 &  69.7 &  17.1 & 55.6 & 5.99   &             \\ 
 lobe 1 (SW)   &  4.46        &  42.76        &  21.14 & 11.63 & 57.1  & 24.0 &           &       &       &       &      &        &             \\
 lobe 2 (NE)   &  5.15        &  50.62        &  19.46 & 13.13 & 45.0  &  29.5&           &       &       &       &      &        &             \\ \hline
\end{tabular}
}
\label{table:RadioComps} 
\\
Note - FIRST flux densities of lobes of J1352+5348 is sum of several small components. `C' : core, `L': lobe, PS : point source. 
Sizes are deconvolved from synthesized beam.
\end{minipage}
\end{table*}

\begin{table*}
\begin{minipage}{180mm}
\caption{Galaxy groups/clusters parameters}
\resizebox{18cm}{!}{
\begin{tabular}{@{}cccccccccccccc@{}}
\hline
             &          &                  &                     & \multicolumn{10}{c} {galaxy group parameters} \\ \cline{5-14}
Source       & redshift & redshift         & N$_{\rm gal}$       & dist   &  RA   & DEC &  z$_{\rm cmb}$ & N$_{\rm gal}$ & R$_{\rm vir}$ & R$_{\rm max}$ & M$_{\rm NFW}$      &  M$_{\rm Her}$      & Density   \\
name         & (z)      & range        &  0.5, 1.0, 2.0, 4.0 Mpc & (kpc)  & (deg) & (deg) &                &               & (Mpc)         & (Mpc)          & (10$^{12}$ M$_{\odot}$) & (10$^{12}$ M$_{\odot}$) & (1Mpc/h)  \\ 
 (1)         & (2)      & (3)          &  (4)                    & (5)    & (6)   & (7)   &     (8)        &     (9)       & (10)          & (11)           & (12)                 & (13)                 & (14)       \\ \hline
J0836+0532   & 0.0994   & 0.0961 - 0.1027  &  2, 4, 4, 7         & 77.6   & 129.22136 & +05.54070 & 0.09978 & 3 & 0.3335 & 0.3228 & 30.5 & 50.8 & 153.328 \\
J1159+5820$\star$  & 0.0537   & 0.0504 - 0.0570  &  1, 2, 7, 13  &        &           &           &         &   &        &        &      &      &        \\
J1352+3126   & 0.0452   & 0.0419 - 0.0485  &  3, 3, 11, 38       & 129.1  & 208.11753 & +31.46362 & 0.04556 & 3 & 0.1470 & 0.1511 & 1.55 & 2.51 & 74.321  \\
J1649+2635   & 0.0545   & 0.0512 - 0.0578  &  9, 21, 38, 64      & 229.4  & 252.35435 & +26.64486 & 0.05448 & 19 &  0.2878 & 0.6936 & 51.5 & 86.2 & 197.75  \\ \hline
\end{tabular}
}
\label{table:Environ} 
\\
Note - 
Column 1: Source name; column 2: redshift; column 3: redshift range spanning $\Delta$z = $\pm$0.0033 {\ie}covering 1000 km s$^{-1}$ velocity 
dispersion centred at the redshift of radio galaxy host; column 4: Number of galaxies (N$_{\rm gal}$) with redshift z${\pm}{\Delta}$z lying in 
concentric circular regions of radii 0.5 Mpc, 1.0 Mpc; 2.0 Mpc and 4.0 Mpc centred at the position of radio galaxy host; 
column 5: distance of the centre of group from host of 
radio galaxy; column 6 and 7: RA, DEC of the centre of the group; column 8: CMB-corrected redshift of the group; 
column 9: Number of galaxies within the group; column 10: Virial radius of the group (R$_{\rm vir}$); column 11: Maximum radius of the group 
(R$_{\rm max}$); column 12: Mass of the galaxy group using NFW profile (M$_{\rm NFW}$); column 13:  Mass of the galaxy group using Hernquist profile (M$_{\rm Her}$); 
column 14: Normalized environmental density (mean of group galaxy densities) of the group for smoothing scale of 1 Mpc. \\
$\star$ : Source has not been associated to any group in \cite{Tempel14} catalogue.

\end{minipage}
\end{table*}

\section{Notes on individual sources}
In this section we discuss the properties of individual spiral-host double-lobe radio galaxies. 

\subsection{J0836+0532}
This spiral-host double-lobe radio galaxy is reported for the first time. Below we describe the optical and radio properties of this source.
\subsubsection{Host galaxy} 
J0836+0532 is a face-on spiral galaxy with two distinct spiral arms seen in the SDSS image 
(see Figure~\ref{fig:SDSSImages}). The Galaxy Zoo catalogue classifies it a spiral galaxy with 95$\%$ probability. 
The surface brightness profile of the $r$-band SDSS image is best fitted with a combination of S\'ersic and exponential components.
The best fitted parameters {\ie}bulge-to-total ratio (B/T) $\sim$ 0.15, bulge radius ($\sim$ 0.64$\pm$0.06$\arcsec$), 
disk radius ($\sim$ 3.82$\pm$0.04$\arcsec$) 
and S\'ersic index (n $\sim$ 0.83$\pm$0.06) suggest it to be a disk dominated galaxy 
(see Table~\ref{table:OptMorph}). 
The apparent projected size of this galaxy is at least $\sim$ 25$\arcsec$, which 
corresponds to $\sim$ 44.0 kpc, therefore indicating it to be a giant spiral galaxy.
The optical colour (u - r = 2.64, see Table~\ref{table:Mags}) and spectral continuum (see Figure~\ref{fig:J0836+0532}) suggest 
it to be a red galaxy. 
Using the criterion `(g-r) $>$ 0.63 - 0.02(M$_{\rm r}$ + 20)' given in \cite{Masters10}, J0836+0532 (g-r = 0.88, M$_{\rm r}$ = -22.91) 
can be classified as a red galaxy. 
Red spiral galaxies have lower SFR than their blue counterparts, but had similar star formation histories until 
$\sim$ 500 Myr ago \citep{Tojeiro13}. 
The mid-IR colour ([3.4] - [4.6] = 0.45) indicates that the mid-IR emission is dominated by star formation rather than AGN. 
The estimated SFR from 22~$\mu$m is $\sim$ 9.99$^{+9.66}_{-3.76}$ M${_{\odot}}$ yr$^{-1}$ and so it is
an actively star forming galaxy.
\par
The SDSS optical spectrum of the central region (3$\arcsec$ in diameter) shows strong permitted 
({\ie}H$_{\beta}$ ${\lambda}$ 4862 {\AA}, H$_{\alpha}$ ${\lambda}$ 6563 {\AA}) and forbidden 
({\ie}[O II] $\lambda$ 3726 {\AA}, [O III] ${\lambda}$ 5007 {\AA}, [N II] ${\lambda}$ 6548 {\AA}, [S II] ${\lambda}$~${\lambda}$ 6717, 6731 {\AA}) 
emission lines (see Figure~\ref{fig:J0836+0532}). The emission line flux ratio diagnostics indicate 
it to be a Seyfert or High Excitation Radio Galaxy (HERG) (see Table~\ref{table:LineRatio}). 
The black hole mass estimated from the bulge magnitude using the black hole mass--bulge luminosity empirical relations is around 
$\sim$ 2.7 $-$ 5.5 $\times$ 10$^{8}$ M$_{\odot}$. 
Thus, the black hole mass in this galaxy is similar to those found in powerful radio-loud AGN \citep[{\eg}][]{Chiaberge11}. 
Only large spirals are known to host such massive black holes. 
The black hole mass in this galaxy is similar to that found in previously reported spiral-host radio galaxies 
({\eg}0313-192 \citep{Keel06}; J2345-0449 \citep[]{Bagchi14}).     
\subsubsection{Radio morphology} 
The NVSS image shows a distinct core-lobe morphology with total end-to-end projected linear size of ${\sim}$ 4$^{\prime}$ that 
corresponds to $\sim$ 420 kpc at the redshift of 0.099. Table~\ref{table:RadioComps} lists flux densities, sizes and 
separations of the different emitting 
components seen in the FIRST and NVSS images. The FIRST image shows a clear core component but barely detectable extended emission from the lobes. 
The radio morphology is clearly detectable only in the lower resolution (beam-size $\sim$ 45$\arcsec$) NVSS image. 
In the NVSS image, the peaks of the South-West (SW) and North-East (NE) lobes lie at 74$\arcsec$.5 and 80$\arcsec$.0, respectively, 
from the core component. 
The flux densities of the core, NE and SW lobes are 23.2 mJy, 13.8 mJy and 25.4 mJy, respectively, indicating that 
the NE lobe is significantly weaker than the SW lobe. 
The distinct lobes with signs of edge brightening in the NVSS image suggest it to be a FR-II radio galaxy. 
However, the total 1.4 GHz NVSS luminosity ($\sim$ 1.53 $\times$ 10$^{24}$ W Hz$^{-1}$) lies below the FR-I/FR-II 
luminosity break. 
A radio galaxy with large-scale FR-II radio morphology but low radio luminosity can be possible if the 
radio-lobes are old {\ie}the source is in the later phase of its lifetime. 
Evolutionary models of radio galaxies suggest a slow increase in both size and luminosity over most of the source lifetime, 
followed by a rapid decline in luminosity at the late phase of its lifetime \citep[{\eg}][]{Luo10}. 
The inflated radio morphology of this source further indicates the possibility of being it at the late phase of its 
evolution, although the presence of bright compact radio core and hotspots implies that the AGN is still active. 
Multi-frequency radio observations are required to estimate the age and evolutionary stage of this radio source. 
Optical emission line ratio diagnostics suggest it to be a HERG (see Figure~\ref{fig:KewleyPlot}). 
In fact, \cite{Best12} also classify this source as a HERG 
with `high-excitation radio-mode' accretion. 
In general, HERGs are believed to possess relatively high accretion rates in radiatively efficient standard accretion discs 
fuelled by cold gas, possibly brought in through mergers and interactions, and with some of the cold gas leading to associated star formation 
\citep[{see}][]{Best12}.
\begin{figure*}
\includegraphics[angle=0,width=7.0cm,trim={2.9cm 0.0cm 0.0cm 0.0cm},clip]{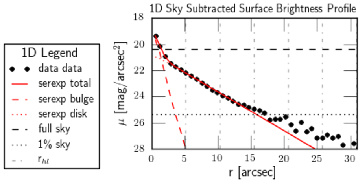}{\includegraphics[angle=0,width=11.0cm,trim={0.5cm 0.7cm 2.8cm 3.0cm},clip]{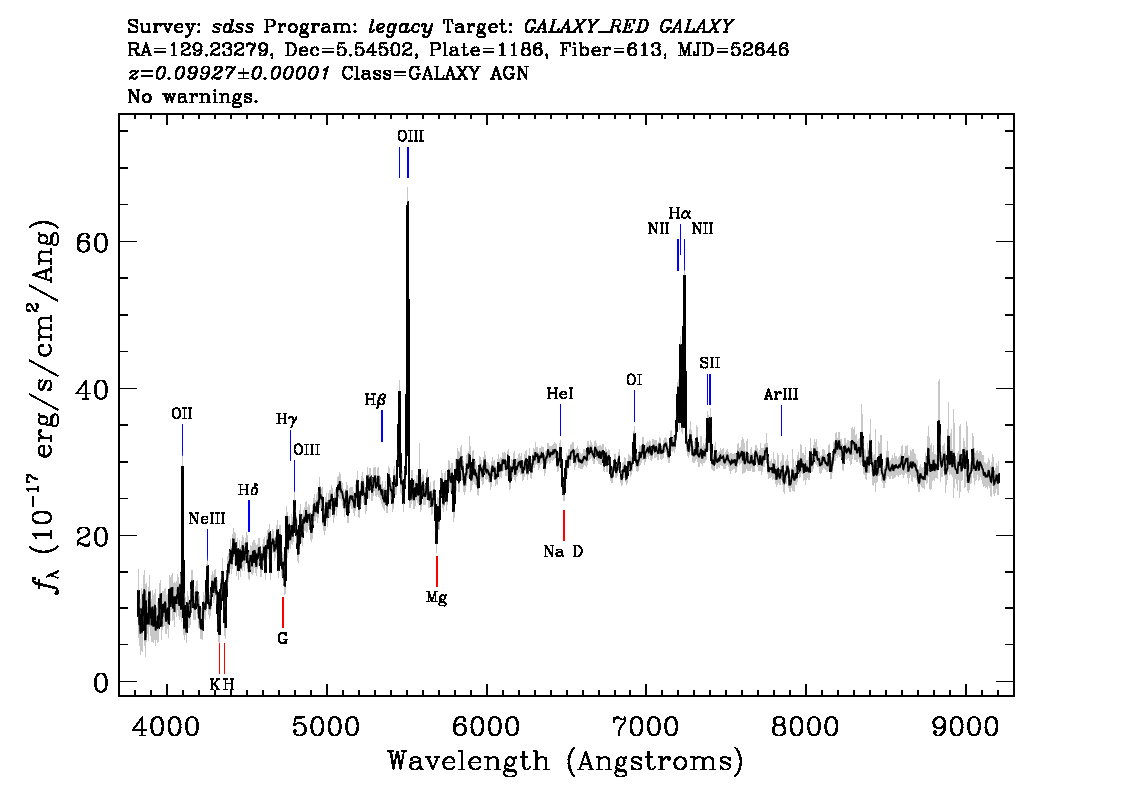}}
\includegraphics[angle=0,width=9.0cm,trim={0 1.75cm 0 0},clip]{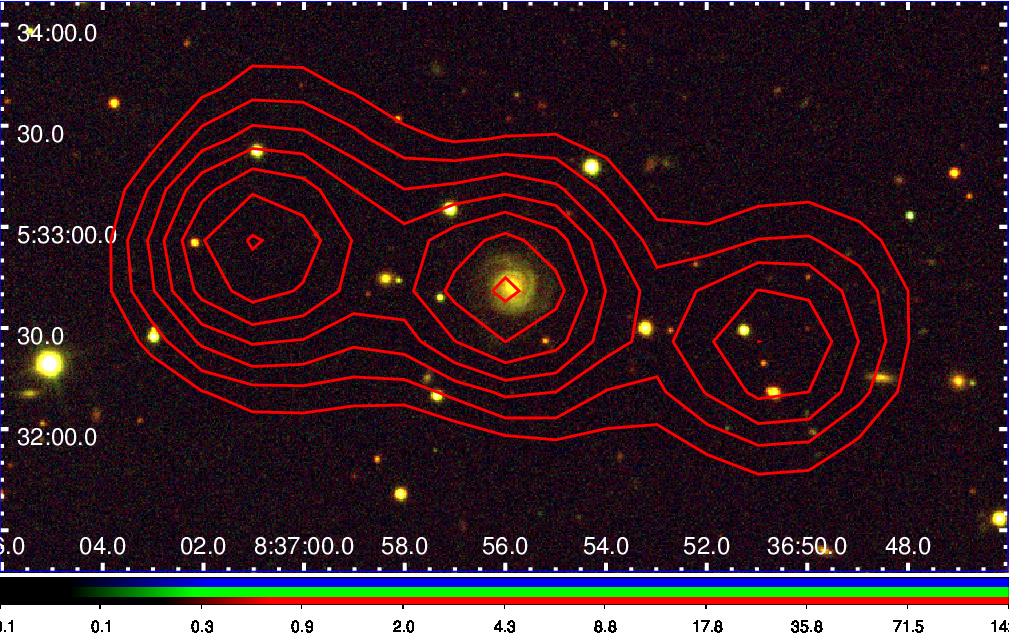}{\includegraphics[angle=0,width=9.0cm,trim={0 1.75cm 0 0},clip]{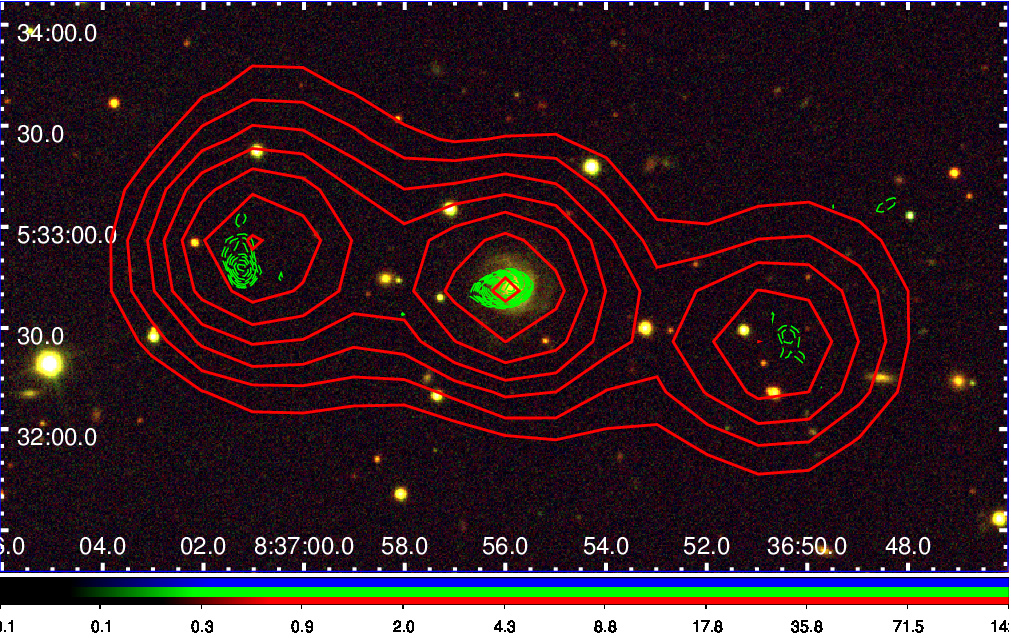}
}   
\caption{J0836+0532: {\it Top left panel:} Best fitted optical surface brightness profile 
(Where black points are data points. Solid red curve, dashed red curve and dotted red curve represent total best fit, bulge component 
and disk component, respectively. Dashed and dotted horizontal lines represent full and 1$\%$ sky, respectively).
{\it Top right panel:} SDSS optical spectrum.   
{\it Bottom left panel:} NVSS contours in solid red overlaid on the SDSS optical image 
(NVSS contour levels are at 2.5, 4.0, 5.8, 7.9, 10.5, 13.5 and 15.0 mJy).  
{\it Bottom right panel:} NVSS contours (in solid red) and FIRST contours (in dashed green) overlaid on the SDSS optical image 
(FIRST contour levels are at 0.6, 0.75, 0.95, 1.2, 1.45, 1.8, 2.1, 2.7, 3.5 and 4.6 mJy). 
In all images North is upward and East is leftward. In contours overplotted on optical images, x-axis and y-axis 
are RA (hour, minute, second) and declination (degree, arcmin, arcsec), respectively. 
Same plotting convention is followed in Figure 3, 4 and 5.} 
\label{fig:J0836+0532} 
\end{figure*}
\subsubsection{Environment}
Using SDSS spectroscopic data we examine the distribution of galaxies around J0836+0532. 
Considering the typical velocity dispersion in cluster of galaxies to be $\Delta$cz $<$ 1000 km s$^{-1}$ {\ie}$\Delta$z$\pm$0.0033, 
we find that there are only 2, 2, 4 and 7 galaxies in the circular regions of the radii of 0.5, 1.0, 2.0 and 4.0 Mpc centred 
at J0836+0532 position. 
Using the \cite{Tempel14} catalogue, we find that J0836+0532 is associated with a group of 3 galaxies 
with magnitude limit (m$_{\rm r}$) $\leq$ 17. 
The estimated virial radius of the group is $\sim$ 333.5 kpc and the mass is in the range of 3.05 $-$ 5.08 $\times$ 10$^{13}$ M$_{\odot}$.  
\subsection{J1159+5820}
This galaxy located at RA = 11h 59m 05.s8, Dec = +58$^{\circ}$ 20$\arcmin$ 35$\arcsec$.5 with redshift (z) = 0.054, and 
has been identified as `CGCG 292$-$057' \citep{Stoughton02}. 
\cite{Koziel12} present a detailed optical and radio study of this source.   
\subsubsection{Host galaxy}
The SDSS image of J1159+5820 shows a disturbed optical morphology in which the bright central bulge is accompanied by spiral arms 
resembling tidal tails (see Figure~\ref{fig:SDSSImages}).  
The optical surface brightness profile is best fitted with a combination of S\'ersic plus exponential components. 
The best fitted parameters suggest it to be a bulge dominated system in which $\sim$ 75$\%$ of the total light is coming from the bulge 
({\ie}B/T $\sim$ 0.75). The total span of this galaxy, including spiral arm tails, is nearly $\sim$ 75$\arcsec$, which corresponds to 77.8 kpc.      
The optical morphology and sizes of the different components indicate it to be a merged system, viewed close to face on (b/a $\sim$ 0.75).
This may be a recent merger of a spiral galaxy and an elliptical galaxy, where the bulge representing elliptical galaxy is nearly intact while 
the spiral galaxy is tidally disrupted.  
The optical spectrum shows strong H$\alpha$, [O II] emission lines and redder continuum that can be understood as the presence of both young and old 
stellar populations. Emission line flux ratio diagnostics suggest it to be a LINER or Low Excitation Radio Galaxy (LERG) 
(see Table~\ref{table:LineRatio}).     
The SFRs derived from the mid-IR and UV emission are 2.89$_{-1.16}^{+2.78}$ M$_{\odot}$ yr$^{-1}$ and 1.50 M$_{\odot}$ yr$^{-1}$, respectively. 
The estimated black hole mass is $\sim$ 2.7 $-$ 5.7 $\times$ 10$^{8}$ M$_{\odot}$.  
\subsubsection{Radio morphology}
J1159+5820 (CGCG 292$-$057) has a very peculiar and complex radio morphology. 
The FIRST image of relatively higher resolution (beam-size $\sim$ 5$\arcsec$.0) shows a core and two lobes with flux densities 
of 2.11, 2.14 and 4.36 mJy, respectively. 
The eastern and western lobes are located at 10$\arcsec$.0 and 12$\arcsec$.5 from the core component. 
The total end-to-end radio size is $\sim$ 28$\arcsec$.7, which corresponds to 
$\sim$ 29.8 kpc and lies within the optical size of the galaxy.  
The NVSS image shows large scale radio emission exhibiting `X-shape' or `Z-like' morphology. 
It also shows an outer pairs of lobes that are nearly coaxial with the inner pair of lobes seen in the FIRST image. 
Both outer lobes are accompanied by wing-like structures of low-surface-brightness that are oriented at an angle of $\sim$ 40$^{\circ}$ {\it w.r.t} the 
jet-lobe axis. 
The outer lobes and wings are nearly symmetric {\it w.r.t.} the host galaxy. 
The total end-to-end linear projected size of the outer lobe structure is $\sim$ 6$\arcmin$.3, which corresponds to $\sim$ 392 kpc,
while the total end-to-end linear projected separation of the wings is $\sim$ 7$\arcmin$.9 (494 kpc). 
In J1159+5820, the presence of two pair of lobes is an evidence for episodic AGN activity, where the inner lobes were
formed by recent AGN activity, while the outer lobes were formed during a previous cycle of AGN activity. 
Such galaxies are known as Double-Double Radio Galaxies (DDRGs) \citep[see][]{Saikia09}.   
The wings seen in J1159+5820 may be the fading lobes of another cycle of AGN activity, and, therefore this source 
can be classified as triple-double radio galaxy where three pairs of lobes are seen from 
three epochs of AGN activity \citep{Koziel12}. 
Such radio galaxies are extremely rare and J1159+5820 may be one among rare sources, with the first such source 
B0925+420 being reported by \cite{Brocksopp07}.  
\begin{figure*}
\includegraphics[angle=0,width=6.5cm,trim={2.9cm 0.0cm 0.0cm 0.0cm},clip]{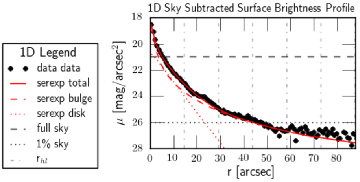}{\includegraphics[angle=0,width=11.5cm,trim={0.2cm 0.5cm 2.5cm 2.9cm},clip]{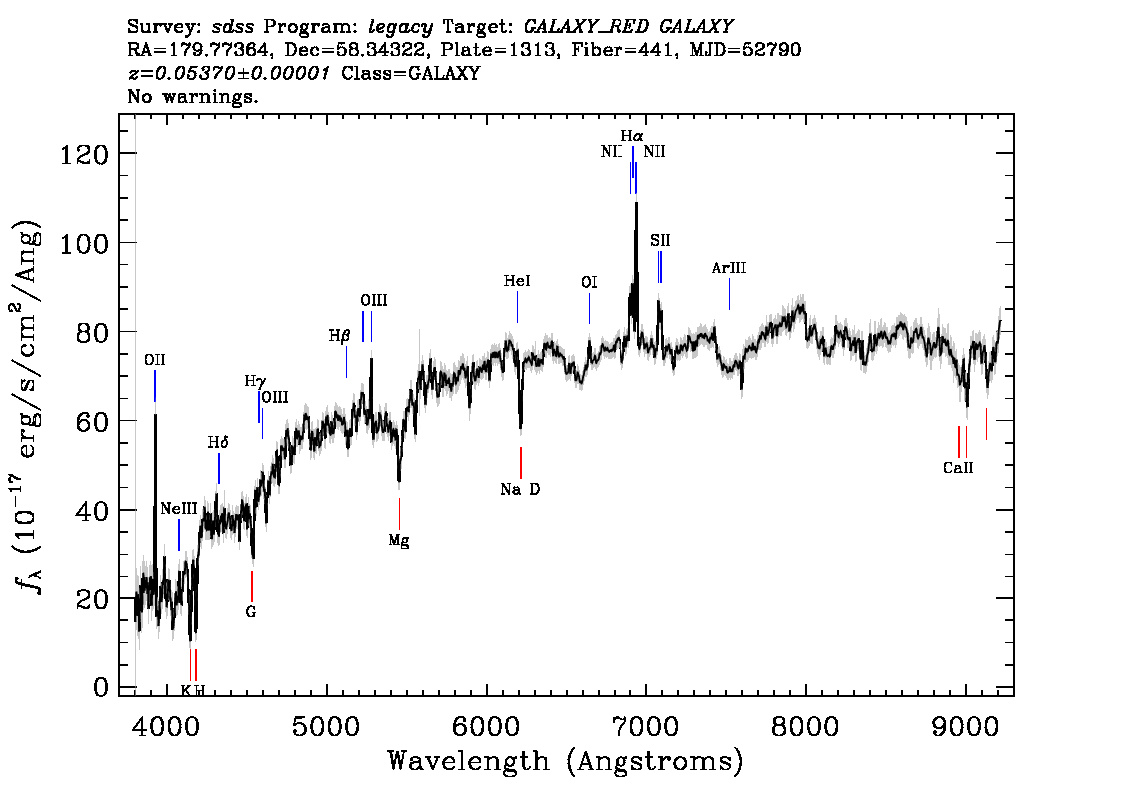}}
\includegraphics[angle=0,width=9.0cm,trim={0 1.75cm 0 0},clip]{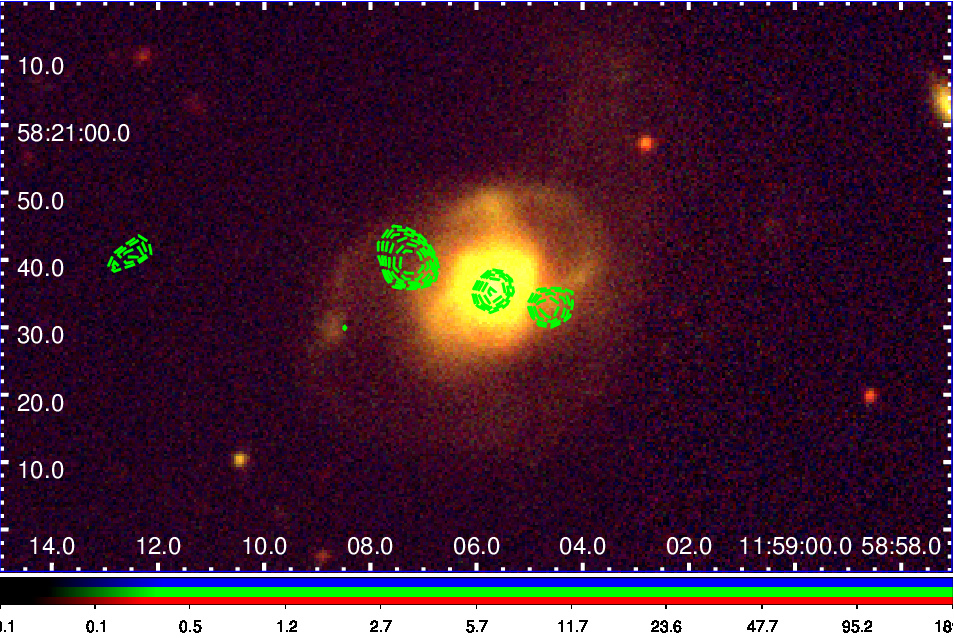}{\includegraphics[angle=0,width=8.85cm,trim={0 1.75cm 0 0},clip]{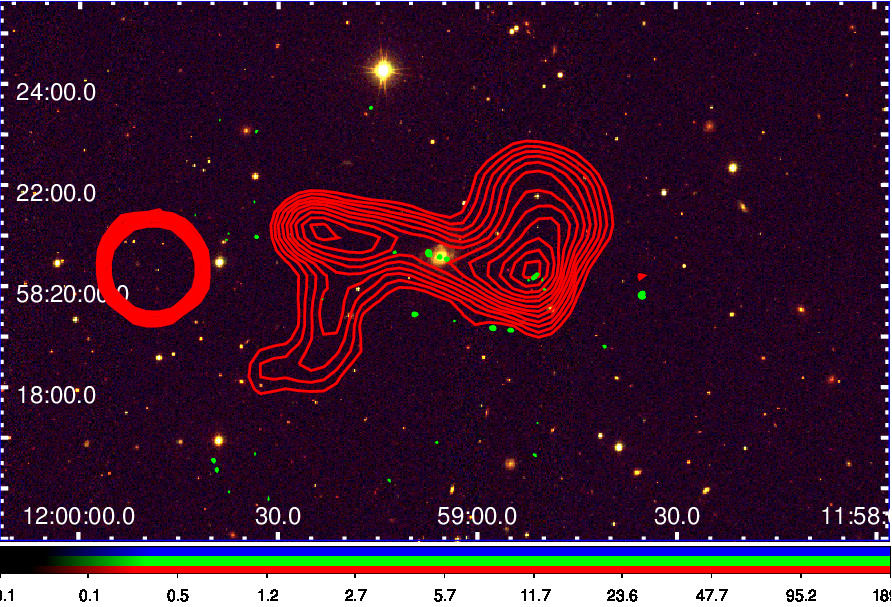}}
\caption{J1159+5820: {\it Top left panel:} The best fitted optical surface brightness profile. 
{\it Top right panel:} SDSS optical spectrum. {\it Bottom left panel:} FIRST contours in dashed green overlaid on the SDSS optical image 
(FIRST contour levels are at 0.75, 9.0, 1.1, 1.3, 1.7 and 2.1 mJy).   
{\it Bottom right panel:} NVSS contours in solid red and FIRST contours in dashed green overlaid on the SDSS optical image 
(NVSS contour levels are at 2.5, 3.0, 4.0, 5.0, 6.5, 8.5, 10.5, 13.5, and 17.5 mJy).}
\label{fig:J1159+5820} 
\end{figure*}

\subsubsection{Environment}
We consider galaxies in the redshift bin of 0.0504--0.0570 and found that there are only 1, 2, 7 and 13 galaxies within 
the circular regions of the radii of 0.5, 1.0, 2.0 and 4.0 Mpc, respectively, centred at this galaxy. 
This suggests the lack of an over-density in the vicinity of J1159+5820. 
In the \cite{Tempel14} catalogue of groups and clusters, this galaxy has not been associated with any group or cluster. 
\subsection{J1352+3126}
J1352+3126 is a well studied source across nearly all wavelengths. 
\subsubsection{Host galaxy}
In the literature, J1352+3126 (other names : 3C 293, UGC 8782 and VV5-33-12  \cite[]{de-Vaucouleurs91}) has been 
classified as both a spiral galaxy \citep[{\eg}][]{Sandage66,Colla75,Burbidge79} and an irregular galaxy 
\citep[{\eg}][]{Tremblay07}, owing to its extremely complex morphology, with a dusty disk, compact knots, and large dust lanes 
(see Figure~\ref{fig:SDSSImages}). 
\cite{Donzelli07} showed that the radial brightness profile can be better fitted by a Sersic law plus an exponential disk component. 
In the Galaxy Zoo catalogue, this galaxy has been classified as a spiral with 86$\%$ probability. 
The optical surface brightness profile is best fitted with a combination of S\'ersic and exponential models. 
The best fitted parameters suggest it to be a bulge dominated (B/T $\sim$ 0.61) galaxy with major-to-minor axis ratio (b/a) $\sim$ 0.60.  
Given its complex morphology and the presence of vast amounts of dust and gas, it is believed that this galaxy is the result of 
a gas-rich merger of an elliptical galaxy and a spiral galaxy \citep{Martel99,deKoff2000,Capetti2000,Floyd06}.
The SDSS optical spectrum shows a dominant red continuum along with strong H$_{\alpha}$ and [O II] emission lines 
(see Figure~\ref{fig:J1352+3126}), indicating the presence of both old and young stellar populations.   
Emission line flux ratio diagnostics suggest it to be a composite galaxy, where both star formation as well as AGN contribute significantly 
(see Table~\ref{table:LineRatio}). 
In fact, detailed stellar population studies suggest a mix of a young (0.1 - 2.5 Gyr) as well as an old (10 Gyr) 
stellar population, with total stellar mass of $\sim$ 2.8 $\times$ 10$^{11}$ M$_{\odot}$ \citep[{\eg}][]{Tadhunter05,Tadhunter11}. 
Hubble Space Telescope (HST) observations have revealed several regions of recent star formation \citep{Baldi08}.
The estimated SFRs from mid-IR and UV continuum are 7.41$^{+7.12}_{-2.99}$ M$_{\odot}$ yr$^{-1}$ and 0.97 M$_{\odot}$ yr$^{-1}$, respectively. 
The low SFR derived from UV measurements is likely due to the presence of dust that causes extinction. 
Using CO line observations \cite{Labiano14} showed the presence of massive molecular gas (M(H2) $\sim$ 2.2 $\times$ 10$^{10}$ M$_{\odot}$) 
distributed along a large $\sim$ 21 kpc diameter warped disk rotating around the AGN and average star formation rate of 
$\sim$ 4.0$\pm$1.5 M$_{\odot}$ yr$^{-1}$. 
The black hole mass derived from the bulge magnitude is $\sim$ 1.3 $-$ 1.9 $\times$ 10$^{8}$ M$_{\odot}$.   
\subsubsection{Radio morphology}
FIRST observations only show a bright central component of 3.7 Jy, with a barely detected north-west lobe and no detection of a south-east lobe. 
The large beam of NVSS helps to detect extended lobe emission where the south-east lobe is less prominent. 
The total extent of this radio galaxy as seen in the NVSS image spans to $\sim$ 335 kpc with total observed 1.4 GHz 
luminosity $\sim$ 1.9 $\times$ 10$^{25}$ W Hz$^{-1}$.
Previous radio observations report this as a FR-II radio galaxy with the two lobes highly asymmetric in intensity and 
a hotspot seen in north-west lobe \citep[{\eg}][]{Bridle81,Beswick04}.
The central component has a steep radio spectrum and high resolution observations of it reveal a compact double-lobed source with 
multiple components and a flat-spectrum radio core \citep{Akujor96,Beswick04,Giovannini05}. 
The projected linear separation of the two inner double lobes is $\sim$ 1.7 kpc and has been interpreted to arise from 
a more recent cycle of AGN activity.
Using multi-frequency radio observations \cite{Joshi11} estimated the spectral ages of the outer and inner lobes to be 
$\leq$ 17 - 23 Myr and $\leq$ 0.1 Myr, respectively.

\subsubsection{Environment}
Considering galaxies within the redshift bin of 0.0419--0.0485 from the SDSS DR10 spectroscopic catalogue, we find only 3, 3, 11 and 38 
galaxies lie in the circular regions of the radii of 0.5, 1.0, 2.0 and 4.0 Mpc, respectively, centred at the position of J1352+3126. 
This suggests that J1352+3126 does not reside in an over-dense region. 
In the \cite{Tempel14} catalogue, this galaxy is part of a group of three galaxies with virial radius of 147 kpc and mass 
$\sim$ 1.55 $-$ 2.51 $\times$ 10$^{12}$ M$_{\odot}$.
\begin{figure*}
\includegraphics[angle=0,width=6.5cm,,trim={2.9cm 0.0cm 0.1cm 0.0cm},clip]{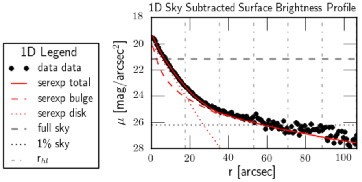}{\includegraphics[angle=0,width=11.0cm,trim={0.1cm 0.5cm 2.5cm 2.9cm},clip]{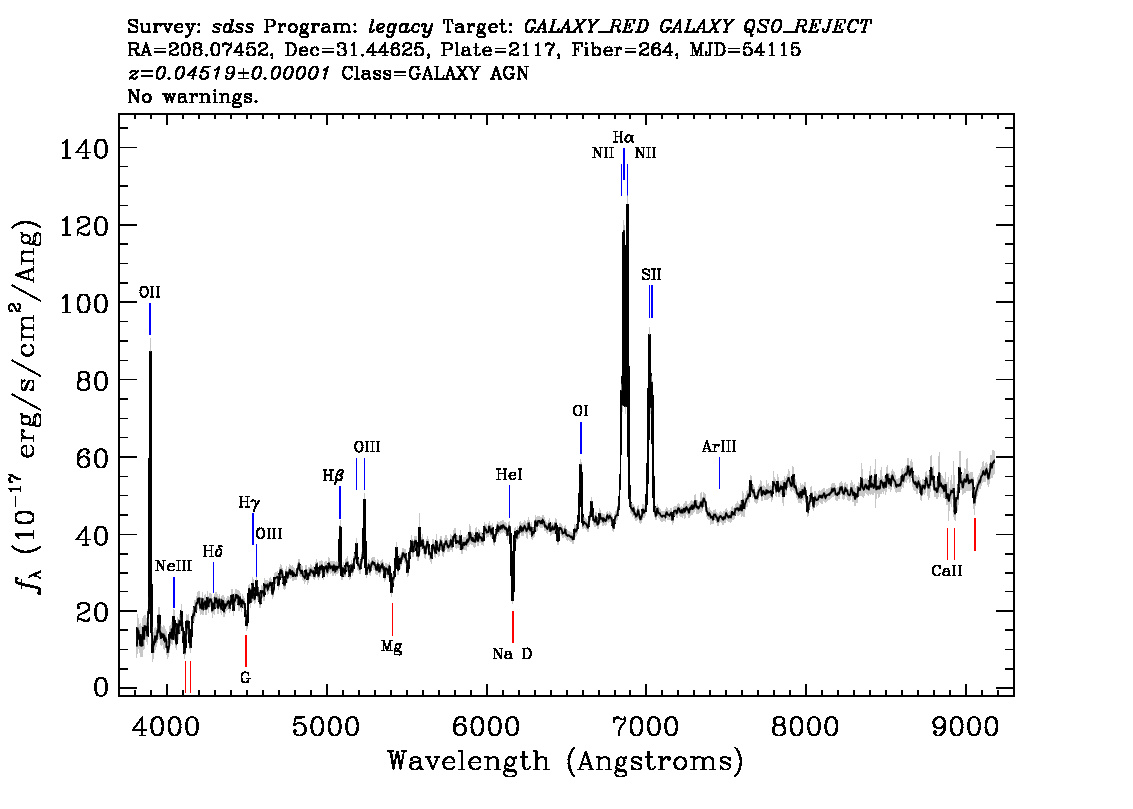}}
\includegraphics[angle=0,width=16.0cm,trim={0.0cm 1.75cm 0.0cm 0.0cm},clip]{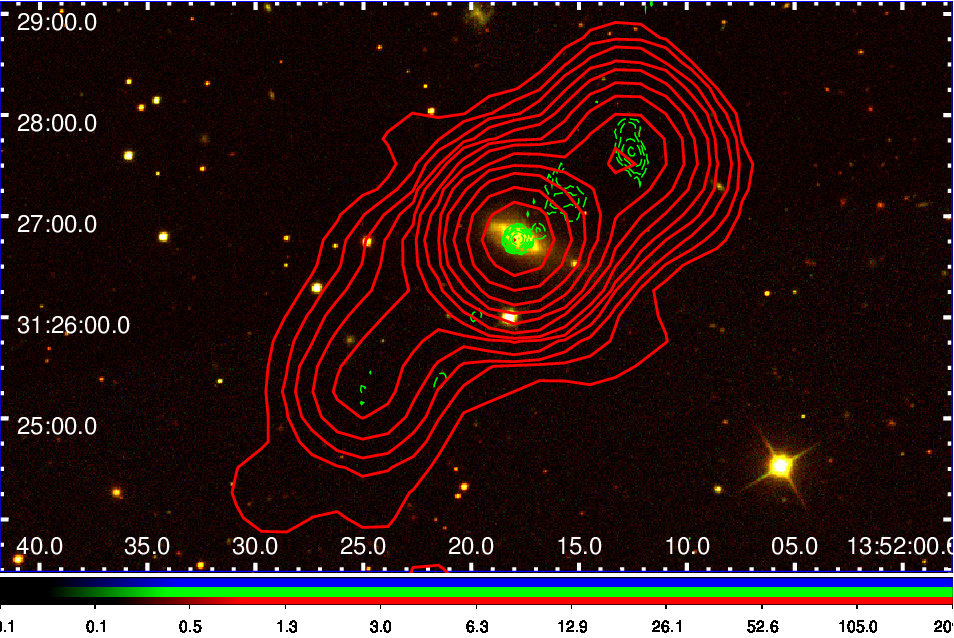}
\caption{J1352+3126: {\it Top left panel:} The best fitted optical surface brightness profile.  
{\it Top right panel:} SDSS optical spectrum.
{\it Bottom panel:} NVSS (in solid red) and FIRST (in dashed green) contours overlaid on SDSS optical image 
(NVSS contour levels are at 2.5, 5.5, 10, 20, 35, 65, 120, 200, 380, 675 and 1200 mJy, and FIRST contour levels are at 
4.0, 6.5, 11.0, 18 and 30 mJy).}
\label{fig:J1352+3126} 
\end{figure*}
\subsection{J1649+2635}
A detailed study of this source has been presented in \cite{Mao15}. 
We independently discovered this spiral-host radio galaxy as one of our sample sources. 
We briefly discuss the properties of this source below.
\subsubsection{Host galaxies}
SDSS image of J1649+2635 clearly shows it to be a face on spiral galaxy with two distinct spiral arms. 
In the Galaxy Zoo catalogue this source has been classified as a spiral galaxy with 97$\%$ probability.     
Its optical surface brightness profile is best fitted with a combination of S\'ersic and exponential models. 
The best fit parameters yield a bulge$-$to$-$total ratio (B/T) $\sim$ 0.63, suggesting the presence of a prominent bulge. 
Interestingly, the optical surface bright profile extends up to $>$ 60$\arcsec$ ($>$ 60 kpc) indicating 
the existence of an extended optical halo that is fitted by the exponential component. 
The optical spectrum is dominated by a red continuum with a clear 4000{\AA} break, a characteristic of early type galaxies. 
The WISE colour ([3.4] - [4.6] $\sim$-0.01) suggests that the mid-IR emission is dominated by star formation. 
The estimated SFRs from mid-IR and UV continuum are $\sim$ 1.67$^{+1.60}_{-0.67}$ M$_{\odot}$ yr$^{-1}$ and 0.20 M$_{\odot}$ yr$^{-1}$, respectively. 
The black hole mass derived from bulge magnitude is in the range of 1.8 $-$ 3.3 $\times$ 10$^{8}$ M$_{\odot}$. 
\subsubsection{Radio morphology}
The FIRST image shows a clear double-lobe radio morphology with two lobes placed nearly symmetrically around the radio core. 
The core, South-Western (SW) and North-Eastern (NE) lobes have integrated flux densities 8.59, 42.76 and 50.62 mJy, respectively. 
The SW and NE radio lobes are located at 24$\arcsec$.0 and 29$\arcsec$.5 from the optical counterpart, respectively (see Table~\ref{table:RadioComps}). 
The total end-to-end projected linear size of the radio structure is $\sim$ 1$\arcmin$.37, corresponding to $\sim$ 86 kpc at 
the redshift (z) 0.055 of the source. 
The NVSS image shows an unresolved but extended structure along the position angle of $\sim$ 55$^{\circ}$.6.
The total NVSS flux density (155.4 mJy) is higher than the total FIRST flux density (102 mJy) and therefore indicates the presence 
of extended radio emission of low-surface-brightness that remains undetected in the FIRST observations.
This source has also been observed at 4.86 MHz (in the GB 6 survey), 408 MHz (in the B2 survey) and 74 MHz 
(in the VLA Low-frequency Sky Survey) with total 
flux densities 63 mJy, 331 mJy and 640 mJy, respectively (see Table~\ref{table:Radioflux}).
\begin{figure*}
\includegraphics[angle=0,width=6.5cm,,trim={2.9cm 0.0cm 0.0cm 0.0cm},clip]{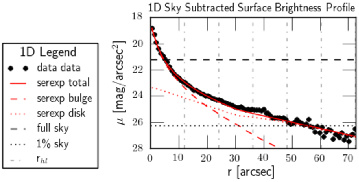}{\includegraphics[angle=0,width=11.5cm,trim={0.5cm 0.6cm 2.5cm 2.9cm},clip]{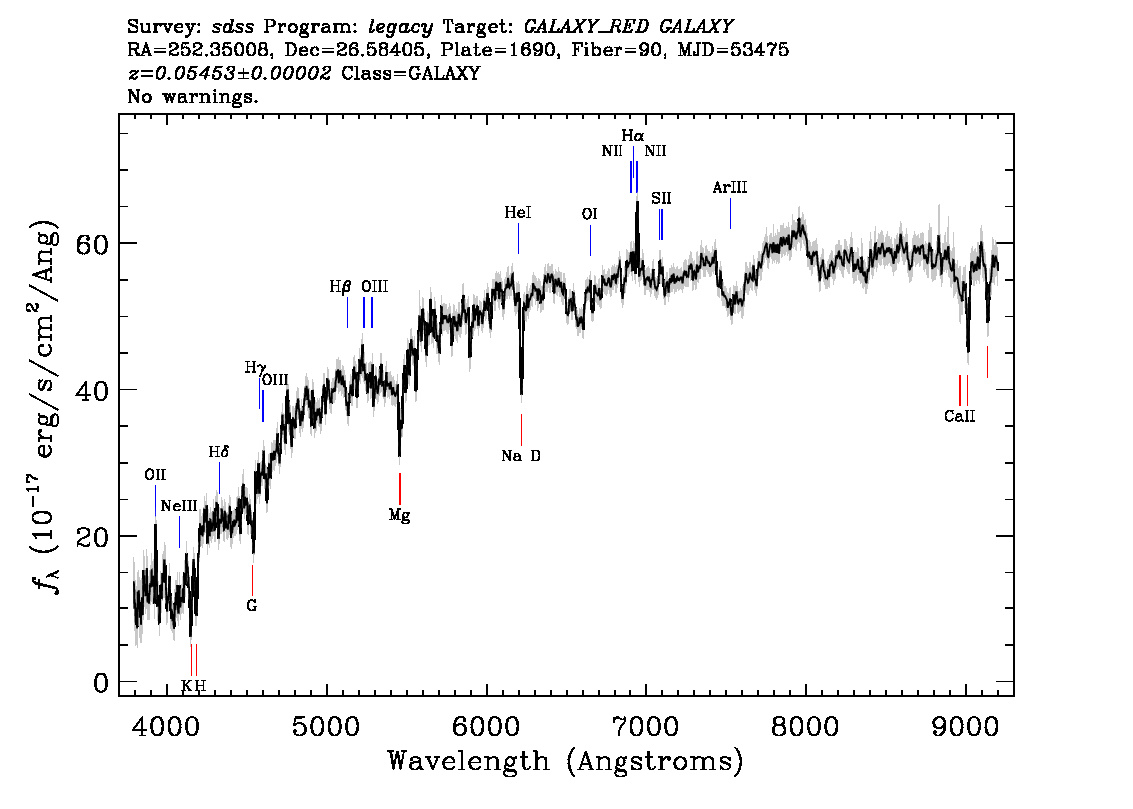}}
\includegraphics[angle=0,width=16.0cm,trim={0.0cm 1.75cm 0.0cm 0.0cm},clip]{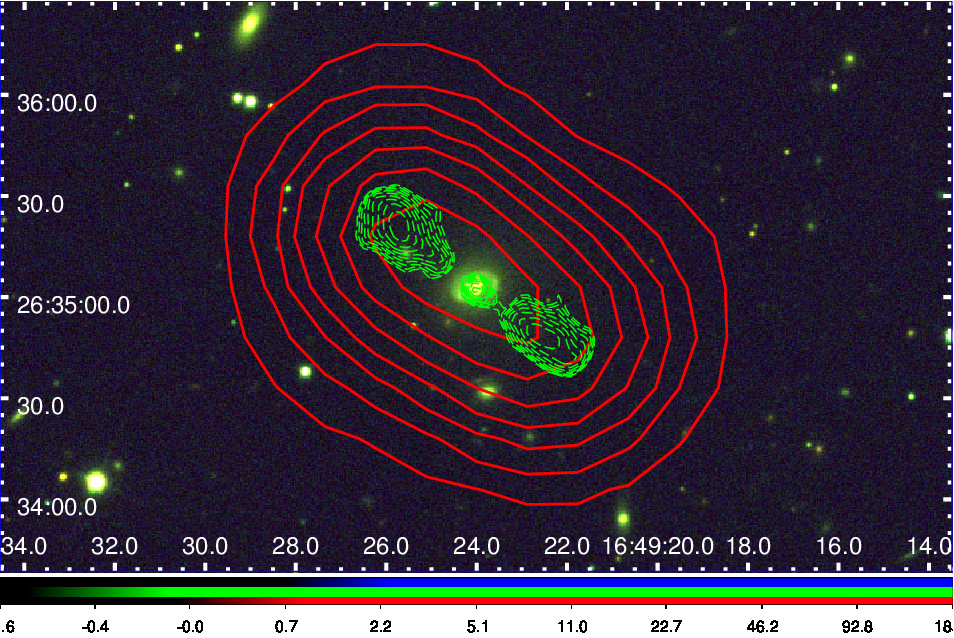}
\caption{J1649+2635: 
{\it Top left panel:} The best fitted optical surface brightness profile. {\it Top right panel:} SDSS optical spectrum.
{\it  Bottom panel:} NVSS (in solid red) and FIRST (in dashed green) contour plots overlaid on SDSS optical image 
(NVSS contour levels are at 2.5, 4.0, 7.0, 11.0, 20.0, 33.0 and 55.6 mJy, and FIRST contour levels are at 1.0, 1.25, 1.55, 2.0, 2.5, 3.0, 4.0 and 5.0 mJy).}
\label{fig:J1649+2635} 
\end{figure*}
\begin{figure*}
\includegraphics[angle=0,width=17.0cm,trim={0.7cm 3.5cm 0.7cm 3.5cm},clip]{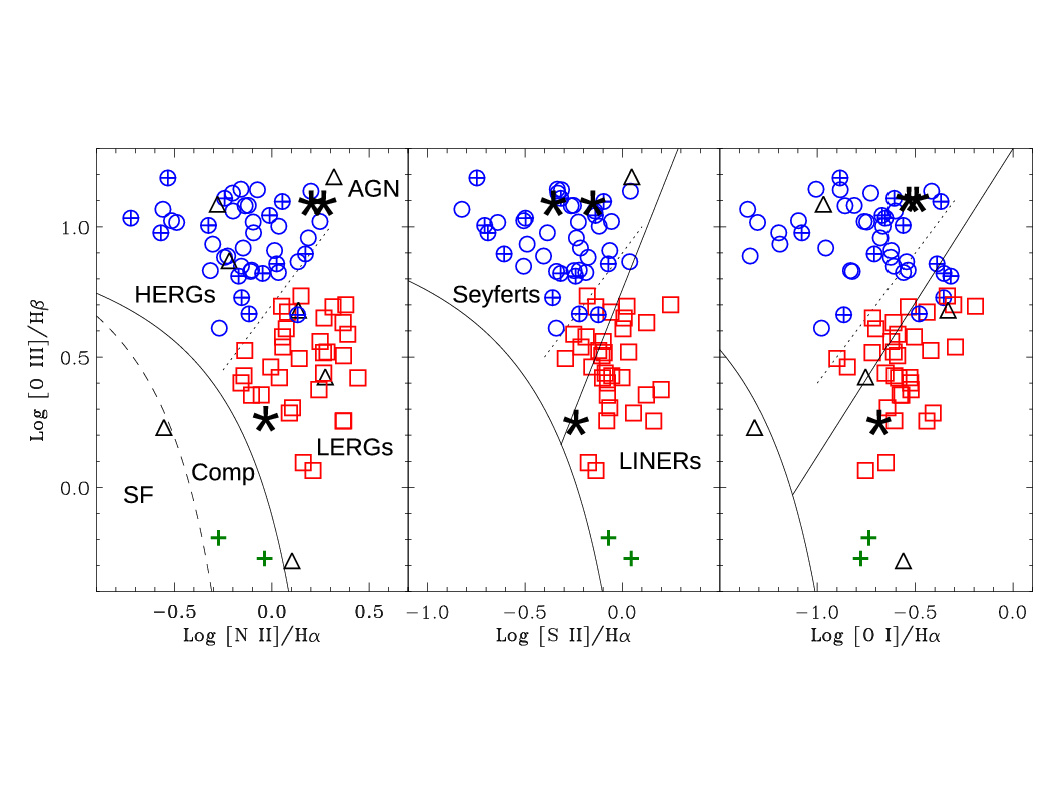}
\caption{Emission line ratio diagnostic plots \protect\citep[{see}][]{Kewley06} for our spiral-host double-lobe radio galaxies overplotted on 
the sample of 3CR radio galaxies from \protect\cite{Buttiglione10}. {\it Left panel} : log [O III]/H$_{\beta}$ versus [N II]/H$_{\alpha}$. 
{\it Middle panel} : log [O III]/H$_{\beta}$ versus [S II]/H$_{\alpha}$. 
{\it Right panel} : log [O III]/H$_{\beta}$ versus [O I]/H$_{\alpha}$. 
The solid curve separates AGN from Star-Forming (SF) galaxies and Composite (Comp) galaxies fall between 
the dashed and solid curves. The `straight solid line' divides Seyfert galaxies (upper left region) from LINERs (right region). 
We note that both radio-quiet (Seyferts, LINERs) and radio-loud sources (radio galaxies) show similar distribution in the emission 
line diagnostic plots \protect\citep[{see}][]{Buttiglione10}. 
The 3CR radio galaxies are classified into High Excitation Radio Galaxies (HERGs in 'blue open circles') and 
Low Excitation Radio Galaxies (LERGs in `red open squares'). 
The dotted lines in each diagram represent approximate boundaries between HERGs and LERGs. 
Our sample sources are in `black stars ($\star$)'. 
The `crossed circles', `green crosses' and `black triangles' represent broad line galaxies, sources with extremely low 
[O III]/H$_{\beta}$, and sources for which the line ratio index can not be estimated, 
as they lack the measurement of one or two diagnostic lines.}
\label{fig:KewleyPlot} 
\end{figure*}
\subsubsection{Environment}
Using SDSS DR10 spectroscopic data and considering galaxies within the redshift bin of 0.0545$\pm$0.0033, 
we find that there are 9, 21, 38 and 64 galaxies in circular regions of radii of 0.5, 1.0, 2.0 and 4.0 Mpc, respectively, 
centred at the position of J1649+2635. This suggests that J1649+2635 is residing in a group or possibly a cluster 
(as more galaxies are likely to be present below the SDSS spectroscopic flux limit). 
In the \cite{Tempel14} catalogue, J1649+2635 is associated with a group of 19 galaxies with viral radius of 287.8 kpc and mass of 
$\sim$ 5.15 $-$ 8.62 $\times$ 10$^{13}$ M$_{\odot}$. 

\section{Discussion}
In our study we find four spiral-host radio galaxies by cross-matching a large sample of 187005 spiral galaxies 
to the FIRST and NVSS catalogues. 
Considering the scarcity of such systems in the local Universe it is important to understand the underlying mechanism(s) 
for the formation of such sources. 
Here we discuss some of the plausible scenarios for the formation of spiral-host double-lobe radio galaxies.  
\subsection{Radio-loud Seyfert galaxies} 
Two of our sample sources (J0836+0532 and J1649+2635) are large face-on spiral galaxies without any 
apparent indication of an ongoing merger or interaction. 
Emission line flux ratio diagnostics suggest J0836+0532 to be a Seyfert or HERG 
(see Table~\ref{table:LineRatio}), as both radio-quiet ({\ie}Seyfert galaxies, LINERs) and radio-loud sources 
({\ie}radio galaxies) can have similar emission line ratios \citep{Buttiglione10}.
In general, Seyfert galaxies represent radio-quiet AGN residing in late type disk dominated spiral or lenticular 
galaxies \citep{Schmidt83,Kellermann89}. 
Moreover, a few examples of Seyfert galaxies with double-lobe radio morphology have been reported 
{\eg}NGC 1052 \citep{Wrobel84}, NGC 1068 \citep{Ulvestad87}, NGC 7674 \citep{Momjian03}, MRK 3 \citep{Kukula99} and 
MRK 6 \citep{Kharb06}). Although, the sizes of radio structures in radio-loud Seyfert galaxies span 
only up to $\sim$ 10--15 kpc. 
The optical properties ({\ie}spiral morphology and emission line ratios) of J0836+0532 and J1649+2635 are 
similar to Seyfert galaxies, but with relatively larger black hole masses 
(2.7 $-$ 5.5 $\times$ 10$^{8}$ M$_{\odot}$ and 1.8 $-$ 3.3 $\times$ 10$^{8}$ M$_{\odot}$, 
respectively). 
Their large black hole masses may play a role in producing powerful jets that result in their large double-lobes 
\citep[{\eg}][]{Chiaberge11}. 
In a recent study, \cite{Kaviraj14} reported that spiral galaxies exhibiting strong nuclear radio emission detected in VLBI observations 
are in general more massive than VLBI-undetected spiral galaxies, suggesting that larger black hole masses result in more powerful jets. 
Moreover, Seyfert galaxies with double-lobe radio morphology spanning few hundreds of kpc are unusual and enigmatic. 
It seems that spiral-host double-lobe radio galaxies constitute a rare population in which optical properties are similar to 
low luminosity AGN ({\ie}Seyferts) while radio properties are of the typical radio galaxies \citep[{\eg}][]{Morganti11}.
\cite{Mao15} proposed that J1649+2635 may have undergone a planar minor merger that would not introduce angular momentum perpendicular to
the plane, so that spiral structure is not disrupted.
The merger would bring sufficient matter (6.6 - 15.5 M$_{\odot}$ yr$^{-1}$ assuming 0.1 efficiency) on to the supermassive black hole and 
trigger radio-loud AGN activity that would eventually result in the large-scale double-lobe radio morphology similar to that seen 
in typical radio galaxies.
This process would also lead to quenching of star formation, making the host galaxy red. 
\subsection{Merger driven radio-loud AGN activity}
Two sources (J1159+5820 and J1352+3126) in our sample are clear cases of mergers between a spiral and an elliptical. 
In both galaxies, the optical morphology and surface brightness profile display the presence of large bulge and 
disturbed disk with tidal tails of spiral arms (see Sections 5.2 and 5.3). 
Both these systems are likely to be recent gas-rich mergers of a massive spiral and an elliptical such that the spiral arm features are disturbed, 
although not destroyed. 
The spiral arms can be sustained over a period of time if there is a large reservoir of gas and dust available, and may perhaps grow
if there is supply of gas and dust from the inner part of the merged system.
In fact, in both these galaxies, the spiral arms appear blue, indicating the presence of young stellar population 
and dust rich local ISM. 
Notably, both these galaxies exhibit episodic AGN activity displayed by the presence of two pairs of lobes. 
It is possible that a large double-lobe radio galaxy may have been residing in a usual early type bulge dominated galaxy and a recent merger 
with a large spiral galaxy brought enough material to the supermassive black hole to give rise a recent episode of AGN activity.          
It has been suggested that the episodic AGN activity is likely to be caused by merger that provides fresh supply of material onto 
SMBH \citep[{\eg}][]{Saikia09,Tremblay10}. 
Recently, \cite{Chiaberge15} demonstrated that galaxy-galaxy merger is the main triggering mechanism for the radio-loud AGN activity in a sample of 
high-redshift (z $>$ 1) sources.
\subsection{Role of environment}  
The large-scale surrounding environment is expected to play an important role in triggering the AGN activity in galaxies. 
It has been suggested that the cooling of hot gas and its inflow towards the centre of the cluster is one of the key 
factors in triggering AGN activity in Bright Cluster Galaxies (BCGs) residing at the centre of clusters 
\citep{McNamara07,Fabian12}.    
It is worth noting that some of the previously found spiral-host radio galaxies ({\eg}J0315-1906 \citep[]{Ledlow98}, 
Speca \citep[]{Hota11}) are reported to be residing in a cluster environment. 
We investigated for the presence of cluster environment around our sample sources using the SDSS DR10 data and the \cite{Tempel14} catalogue, 
and found that none of our sources except J1649+2635 reside in cluster environment. 
J1649+2635 likely resides in a group of galaxies of viral radius $\sim$ 287 kpc and 
dynamical mass of 5.15 $-$ 8.62 $\times$ 10$^{13}$ M$_{\odot}$.   
None of our sample sources is detected in {\em RoSAT} All Sky Survey. 
Therefore, we surmise that the large-scale surrounding environment is not the only possible factor 
in triggering radio-loud AGN activity in spiral galaxies.  
As discussed earlier in this section, galaxy-galaxy mergers 
can also be plausible scenario for the trigger of AGN activity in our sample sources.   
The proposed scenario is consistent with recent works that suggest a strong connection between merging and the onset of 
AGN activity in low-redshift sources \citep[{\eg}][]{Koss10,Ellison11,Kaviraj15}.
\section{Conclusions}
We report the discovery of four spiral-host radio galaxies (J0836+0532, J1159+5820, J1352+3126 and J1649+2635) 
by cross-matching a large sample of 187005 spiral galaxies to 
the FIRST and NVSS catalogues. J0836+0532, with the largest radio size in our sample, is reported for the first time. 
Given our small sample size, despite the use of extremely large optical and radio catalogues, 
it is evident that spiral-host double-lobe radio galaxies are extremely rare. 
In fact, our study can be considered as a test bed to find rare populations of spiral-host double-lobe radio galaxies 
using more sensitive surveys from upcoming facilities such as the Large Synoptic Survey Telescope (LSST; \cite{Ivezic08}) and 
the Square Kilometer Array (SKA; \cite{Dewdney09}).
\\
We note that all of our sample sources exhibit radio morphologies similar to FR-II radio galaxies. 
Since both radio-quiet and radio-loud AGN exhibit similar emission line ratios and therefore, 
diagnostics based on emission line ratios indicate two sources (J0836+0532 and J1159+5820) to be HERG or Seyferts 
and one source (J1352+3126) to be LERG or LINER. 
Considering only optical properties ({\ie}spiral disk and emission line ratios) these sources may have been 
classified as Seyfert and LINER galaxies. 
While radio properties classify them as typical radio galaxies. 
The host galaxies of our sample sources are forming stars at an average rate of $\sim$ 1.7 M$_{\odot}$ yr$^{-1}$ to 10 M$_{\odot}$ yr$^{-1}$. 
The supermassive black hole masses in our spiral-host radio galaxies are a few times 10$^{8}$ M$_{\odot}$, typical of radio-loud AGN. 
Their radio morphologies, with total linear projected sizes ranging from 86 kpc to 420 kpc, are typical of those seen in 
FR-II radio galaxies. However, their NVSS 1.4 GHz radio luminosities are relatively lower, in the range 10$^{24}$ to 10$^{25}$ W Hz$^{-1}$. 
The relatively lower luminosity of FR-II type radio galaxies can be understood if the extended radio emission is old 
{\ie}the electrons are not re-accelerated and the lobes have become dimmer. 
Evolutionary models of radio galaxies generally predict a slow increase in both size and luminosity over most of the source lifetime, 
followed by a rapid decline in luminosity at the late phase of its lifetime \citep[{\eg}][]{Luo10}. 
Therefore, our spiral-host double-lobe radio galaxies may be in the late phase of their evolution.   
Indeed, multi-frequency radio observations of J1159+5820 and J1352+3126 have shown that the extended radio emission in 
these sources is dominated of relic radio emission from the previous phase of AGN activity. 
Our ongoing multi-frequency, multi-resolution radio observations (from Giant Metrewave Radio Telescope) 
of the spiral-host double-lobe radio sources presented in this study would provide further insights on the formation of these systems. \par
Based on properties of the spiral-host radio galaxies in our small sample, 
we propose following plausible scenarios that may be responsible for their formation. \\ 
(i) A radio-loud Seyfert nucleus hosted in a massive spiral galaxy with 
large black hole mass (few times of 10$^{8}$ M$_{\odot}$), similar to radio-loud AGN, may form a spiral-host radio galaxy. 
An AGN with a large black hole mass is likely to produce powerful jets that can lead to the formation of a large double-lobe radio morphology. 
J0836+0532 and J1649+2635 are possible cases without an apparent sign of merger or strong interaction visible in the SDSS images. 
\\
(ii) The optical morphology of J1159+5820 and J1352+3126 suggest them to be recent mergers of 
possibly equally massive spiral and elliptical galaxies. 
In these sources, the spiral structures are disturbed but not destroyed, perhaps due to the presence of large reservoirs of gas and dust. 
Both these galaxies show blue coloured disks and spiral arms, suggesting the presence of young stellar populations and dust-rich local ISM.   
Notably, both these galaxies shows episodic AGN activity, indicating that merger may have resulted in a second episode of AGN activity. 
\\            
None of our sample sources except J1649+2635 are found to be residing in cluster environments,
therefore, suggesting the plausibility of processes, other than accretion through cooling flows, 
in triggering radio-loud AGN activity in our spiral-host radio galaxies. 
We suggest that the large scale double-lobe radio structures found in spiral galaxies are possibly attributed to more than 
one factor, such as the occurrence 
of unusually large supermassive black holes, and/or mergers that do not destroy spiral structures.          

\section*{Acknowledgements}
The financial assistance of the South African SKA Project (SKA SA) towards this research is hereby acknowledged.
Opinions expressed and conclusions arrived at are those of the authors and are not necessarily to be attributed to the SKA SA.  
YW and AB acknowledge support from the Indo-French Centre for the Promotion of Advanced Research (Centre Franco-Indien pour la
Promotion de la Recherche Avanc\'e) under program no. 4404-3. 
We thank the anonymous referee for useful comments that helped to improve the manuscript.
This research has made use of the NASA/IPAC Extragalactic Database (NED) which is operated by the Jet Propulsion Laboratory, California 
Institute of Technology, under contract with the National Aeronautics and Space Administration. 
This publication makes use of data products from the Wide-field Infrared Survey Explorer, which is a joint project of the University of California, 
Los Angeles, and the Jet Propulsion Laboratory/California Institute of Technology, funded by the National Aeronautics and Space Administration.
\\
Funding for SDSS-III has been provided by the Alfred P. Sloan Foundation, the Participating Institutions, the National Science Foundation, 
and the U.S. Department of Energy Office of Science. The SDSS-III web site is http://www.sdss3.org/.
SDSS-III is managed by the Astrophysical Research Consortium for the Participating Institutions of the SDSS-III Collaboration 
including the University of Arizona, the Brazilian Participation Group, Brookhaven National Laboratory, Carnegie Mellon University, 
University of Florida, the French Participation Group, the German Participation Group, Harvard University, the Instituto de Astrofisica de Canarias, 
the Michigan State/Notre Dame/JINA Participation Group, Johns Hopkins University, Lawrence Berkeley National Laboratory, 
Max Planck Institute for Astrophysics, Max Planck Institute for Extraterrestrial Physics, New Mexico State University, 
New York University, Ohio State University, Pennsylvania State University, University of Portsmouth, Princeton University, 
the Spanish Participation Group, University of Tokyo, University of Utah, Vanderbilt University, University of Virginia, 
University of Washington, and Yale University.

\bibliographystyle{mn2e}
\bibliography{firstnvsspaper}

\appendix

\bsp

\label{lastpage}

\end{document}